\documentclass{article}

\usepackage[english]{babel}
\usepackage[usenames]{color}

\begin{document}

\title{Outsourcing Competence%
\thanks{This work has been performed  in the context of Symbiosis, an NWO funded project (NWO is The Netherlands Organisation for Scientific Research), which focuses on software process outsourcing. The title is ambiguous, it can be disambiguated as follows: $X$-competence where ``$X=$ outsourcing''.  When making reference to
this paper we will refer to the authors as J.A. Bergstra, G.P.A.J. Delen, and S.F.M. van Vlijmen. Here
Jan abbreviates Johannes and Bas abbreviates Sebastiaan  as is customary in Dutch.}
}
\author{%
  Jan Bergstra,$^{1,3}$
  Guus Delen,$^{2}$ \&
  Bas van Vlijmen$^{1}$
\\[2ex]
{\small\begin{tabular}{l}
  ${}^1$ Section Theory of Computer Science,
  Informatics Institute, \\
  Faculty of Science,
  University of Amsterdam, The Netherlands.\\
  ${}^2$ Verdonck, Klooster Associates, Zoetermeer, The Netherlands.\\
  ${}^3$ Department of Computer Science,
 Swansea  University, UK.
  \end{tabular}
}
\date{}
}

\maketitle

\begin{abstract}
\noindent  The topic of this paper, competences needed for outsourcing, is organized by first providing a generic competence scheme, which is subsequently instantiated
to the area of sourcing and outsourcing. Sourcing and outsourcing are positioned as different areas of activity,
neither one of which is subsumed under the other one.
 It is argued that  competences relevant for outsourcing
are mainly community based
rather than evidence based. Subjective ability and objective ability are distinguished as categories, together making up ability, which are distinct but not necessarily disjoint from competence.
Conjectural ability is introduced as a form of subjective ability. A person's competence profile includes
competences as well as abilities, including subjective ones.
Competence assessment and acquisition as well as the impact of assessed competence on
practical work is described. The analysis of competence and ability  thus developed  is used as standpoint from
which to extract a specification of an audience for a theory of outsourcing, yet to be written. Moreover,
it allows  to formulate requirements for and in preparation of the development of an outsourcing theory.
Formulating these requirements is done under the assumption that a person's awareness of a
theory of outsourcing is expected to strengthen that person's outsourcing competence profile.
\end{abstract}

\section{Introduction}\label{sec:Intro}

The purpose of this paper is to provide requirements for a theory of outsourcing.%
\footnote{This work aims to contribute to the area of  software process outsourcing,
which is a part of IT-sourcing as well as of software engineering.
An outsourcing theory relevant for software
engineering can be understood as a specialization of a more general  theory of outsourcing,
which, however, we could not retrieve in a satisfactory way from the existing outsourcing  literature.
Now developing a theory of outsourcing from scratch is problematic because it is
dependent on one's objectives and formulating these objectives requires a jargon which the theory is
meant to explain in the first place.

Outsourcing competence turns out to be a notion which can be used in advance of the
development of a theory of outsourcing and which is helpful for preparing a stage where objectives of
and for a theory of outsourcing can be specified. The rationale for classifying this work under software engineering is indirect. It contributes to the requirements engineering that precedes the development of a general outsourcing theory, needed itself for developing a software engineering specific theory of outsourcing.%
}
In particular we will consider outsourcing as a competence%
\footnote{In \cite{MPBR2002} a competence is defined as the ability to do something. Wikipedia
distinguishes several fields of competence, in particular HRM and describes it in that particular context
as  ``a standardized requirement for an individual to properly perform a specific job'';
Merriam-Webster.com defines
competence amongst other uses as the state of being competent, where out of four different options
``having requisite or adequate ability or quality'' is most suited to the objectives of this paper.
BusinessDictionary.com has:
\begin{quote}``A cluster of related abilities, commitments, knowledge, and
skills that enable a person (or an organization) to act effectively in a job or situation.
Competence indicates sufficiency of knowledge and skills that enable someone to act in a wide variety of situations. Because each level of responsibility has its own requirements, competence can occur in any period of a person's life or at any stage of his or her career.''
\end{quote}
 as the first description of competence, besides
a legal meaning which occurs in all dictionaries as well. Further
Merriam-Webster.com explains ability trivially as ``the quality or state of being able'' and also as
``competence in doing'', with ``natural aptitude or acquired proficiency'' as a third meaning.}
which can be acquired and productively exploited. Inspection of the literature on sourcing and
outsourcing demonstrates that this is a high impact area of practice. Even without having any
detailed theoretical framework on outsourcing available, and without having stable working
definitions of the notions involved at hand,  it is possible to analyze
competences relevant for sourcing (and in particular IT sourcing) and outsourcing. Our objective is
to discuss competences relevant for sourcing and for outsourcing in a top-down fashion, in such a way
that one may impose requirements, or at least expectations about what theoretical work on outsourcing
may add to the cluster of competences at hand.

This paper complements our work in  \cite{BV2010} where it is argued that in the definitions of outsourcing and related concepts used and surveyed by Delen in   \cite{Delen2005,Delen2007}%
\footnote{%
In Delen \cite{Delen2005,Delen2007} a survey is given of definitions of outsourcing, insourcing, outtasking, intasking, follow-up sourcing, back-sourcing, greenfield outsourcing and greenfield insourcing. As early references concerning sourcing Delen mentions: \cite{KernWillcocks1999}, \cite{Lacity1993}, \cite{LohVenkatraman1992}. In addition to these notions multiple outsourcing has become prominent, a definition thereof can be found in \cite{Beulen2000} where multiple outsourcing is subsumed under outsourcing.}
and in many later definitions an ambiguity is somehow built in.
An attempt was made in \cite{BV2010} to disambiguate the concepts of insourcing and outsourcing. The key design decision which we will import from that paper is that outsourcing and insourcing  will refer to transformations rather than to steady states arrived at after an appropriate transformation. This view was extended in \cite{BDV2011a} where the term sourcement has been coined for referring to the temporally static assignment of units to
each of the sources in a group of sources, in combination with an assignment to each source of one or more
business processes and themes as being enabled by the source's presence. We will
assume this interpretation of outsourcing and the notion of a sourcement below. A basic sourcement constitutes of a single source, and a link to a unit (its owner), or it constitutes a group of sources each
assigned to the same unit (owner), such that future outsourcing of some but not all sources in the group or outsourcing  of different sources in the group to different units is implausible.

Sourcing is then conceived as the provisioning of sourcements, whereas outsourcing, insourcing, backsourcing, and follow-up outsourcing denote classes of transformations of sourcements seen
from the perspective of a specific unit and its mission.

\subsection{Terminology}
Many related terms will be used in the paper, and in the absence of very clear semantic intuitions about the use of these terms we will try to use the in a systematic fashion and explain how that is intended. Unfortunately our explanation of
these conventions below will not unambiguously resolve in all cases which term to use in a specific context.

\subsubsection{Domain, theme, subject, topic, field, area, and aspect}
The terms domain,  theme, subject, topic, aspect, field, and area will be used with the following conventions in mind. A domain is a field of human operation and activity. IT is an example of a domain. A subject is a focus of study,
with outsourcing as an example. At the same time IT can be a subject of study and research.

A topic is a coherent part of a subject, in other words a topic is a sub-subject. Topics have aspects. EU procurement of services is a topic in the context of the subject of service procurement which includes service acquisition by independent organizations as well as by state controlled organizations. Different patterns of EU procurement procurement processes
each constitute aspects of that topic. A theme is an internal or external  business objective of a unit. HRM is an example of a theme in some unit. Of course HRM can also occur as a subject of study and it may be even
be understood as a domain, when undone from its unit specific aspects.

The term area will be used for clusters of subjects as well as for vaguely specified subsets of subjects.
Sourcing and outsourcing is an area. IT outsourcing may also be labeled an area. A field is similar to an area
but with a bias towards demarcation of competence and human activity. As an example: outsourcing consultancy
is a field of professional activity.

It has proven difficult to make use of these terms in a completely systematic way and moreover the mentioned
conventions are not generally agreed upon. Nevertheless we have tried to adhere to this interpretation of these terms even in cases where other use of words seems somewhat more plausible. Many nouns can occur in different roles,
and even a noun may occur in different roles simultaneously in the same occurrence in the same sentence.

As an example for role multiplicity consider $M_w$ for weight measurement.
$M_w$ is a very general area which comprises many specialized competences.
$M_w$ may be a subject of research and teaching at some technical university,
$M_w$ may appear as a theme in a food production chain. Then the theme $M_w$ may for instance be considered
a costly part of the production process.
Domain specific $M_w$ is a field of expertise for workers in some area.
$M_w$ may feature as an aspect of quality control in a plant,
$M_w$ with high precision is a topic in $M_w$, dealing with  the problems that arise if dangerous substances must be measured is an aspect of $M_w$.

\subsubsection{Competence, competency, ability, capability, capacity, skill}
We understand competency and competence to have  the same meaning and of these two we will use competence throughout the paper. Competence and ability are categories which can be instantiated for a domain or an area. Per area competence and ability may range from being identical to being disjoint and both possibilities in between. In outline, but not in all specific contexts, a competence is marked by it being recognized by others as a means to some end, and an ability is marked by it being effective for reaching some goal.

We will make no use of the term capability, but we will assume that capabilities are for groups of persons what
abilities are for individuals. We make no use of the term skill but we will assume that a skill is to be placed between an ability and a competence, being technically well specified like most competences and being oriented towards an objective though not necessarily proven to be effective.  Capacity is often used with the same meaning as capability but it has an
an important additional meaning referring to a type of a role.

\subsection{A pragmatic perspective on outsourcing competence?}
As an introduction to the detailed problem statement of this paper as given in \ref{Survey}
we list a number of viewpoints which together
may constitute a (seemingly) pragmatic perspective on sourcing competence and outsourcing competence. Pragmatic as
used in the title of this Paragraph
is meant to indicate that all words are used with a seemingly most useful meaning and that a story fitting a presentation with a few bullets is sought. The question mark represents doubts concerning this so-called
pragmatic perspective. This critique will be formulated in detail and accordingly redesigning an explanation of outsourcing competence so that this critique is remedied is the core of the paper. (Outsourcing) competence will be understood as a component of  a so-called (outsourcing) competence profile.%
\footnote{Much of this discussion is quite general and independent of the specific subject of sourcing and outsourcing. We failed to find in the literature an explanation of the concepts of competence and ability
with the kind of features that we need. In particular we found no source for the notion of ``conjectural ability'', or
an equivalent notion, that
we will need to position the extension that awareness of novel theory may effect in relation to a person's existing competence. Concerning that extension two conceptual problems are encountered which need to be remedied. The major difficulty is that this extension may potentially lie outside competence in categorical terms. So we will need a bigger container in which to include both competence and this extension. That container is termed ``competence profile''. The minor difficulty is that awareness of novel and perhaps speculative theory need not imply actual ability.
That issue will be settled by speaking of conjectural ability.}

With S\&O we denote the combined area of sourcing and outsourcing.
The propositions below together outline the pragmatic perspective on S\&O competence.

\begin{enumerate}
\item S and O need not be distinguished. Sourcing may represent all of S\&O.
\item S\&O competence results from having been active in one or more roles in practical outsourcing processes.
\item S\&O competence is strengthened by reading generally approved literature about the subject.
\item Adequate certification guarantees adequate competence.
\item A person $P$'s increased S\&O competence leads to an increased probability to succeed in performing
roles in S\&O processes when $P$ is asked to do so.
\item In other words:  competence must lead to ability, because ability is what makes a project succeed.
\item In the area of S\&O, competence and ability need not be distinguisthed. If one insists  on
maintaining that distinction nevertheless, ability is  an abstract functional version of competence.
More generally: $X$ ability is what $X$ competent
persons have if one abstracts from  their past and experience. Competence can be certified by looking at files.
Ability is a property that can be tested and observed.
\item Ability is situation dependent: a competent person may at some moment be unable to perform, but an incompetent person is unlikely to perform adequately in a demanding situation. Avoidable
human mistakes may find their cause either
in situational inability or in structural incompetence. Such issues must be counteracted by different means.
\item Empirical research on S\&O leads to a stronger evidence base for methods and practices that are applied by
persons with S\&O competence.
\item Working towards evidence based S\&O competence is an objective of the S\&O industry. Research
leads to that state. For that reason it must be supported.
\item Only if research adds to the evidence base of S\&O, a person's  awareness of its outcome
effects a growth of that person's competence.
\end{enumerate}

\subsubsection{Critique on the pragmatic perspective}
We will explain in detail why we will distinguish sourcing from outsourcing in the paper. More importantly, however, against this  ``pragmatic theory''  the following arguments can be put forward both in general
and also on the specific case of S\&O.
\begin{itemize}
\item The assumption that competence implies ability presupposes an evidence base for competence.
In the area of S\&O providing that evidence base is currently unachievable.
For that reason a notion of competence is needed that does not imply
ability and that is independent of the availability of an evidence base.%
\footnote{Our work can be criticized for portraying the presence of an evidence base as a matter of yes or no,
that is too much a black and white picture, whereas in practice the presence of an evidence base is a gradual matter.
Taking a degree of being evidence based into account at this stage introduces a formidable additional complexity.
That the analysis of competence and ability can be performed at all on the basis of a black and white scheme is an
assumption that underlies our work.}
\item S\&O ability is a relevant concept if only because it can be hypothesized in spite of
the fact that its assessment requires a
grip on an evidence base which is not remotely available at present. For that reason we will propose to think in terms of conjectural abilities. Such conjectures may be put forward by small communities not representing even a significant minority of the community that decides on competence.
\item Ability is not an abstract version of competence. Conjectural ability may be but need not be a component of competence.
\end{itemize}

\subsubsection{A comparison with evidence based medicine}
In order to explain the conceptual difficulty with the ``pragmatic theory'' let us consider the now increasingly common concept of evidence based medicine (EBM).%
\footnote{The Wikipedia page on alternative medicine is very informative about these matters.} If EBM is a step forward then medical competence has been non-evidence based in the past. The very focus on EBM indicates that non-evidence based medicine is the
starting point. Some hold that conventional medicine is EBM by definition.%
\footnote{This is what the pragmatic theory of competence and ability would imply in the case of medicine. To see this implication it must be assumed that medical actors are competent and that their competence implies ability, which
shows up in an increased success rate for therapeutic actions.} That is unconvincing: conventional medicine is based on a mainstream accreditation of competences (we will denote it with CCBM: conventional competence based medicine below), whereas alternative medicine lies
outside that area. The available evidence base is not sufficiently large to ensure that all actions performed by
medical professionals with conventional competence is evidence based. That is even a theoretical impossibility,
there will always be marginal cases where the evidence collected thus far is insufficient to justify a preference for
selecting a particular course of action, while the need ``to do something'' cannot be denied.
Conventional competence  provides rules of conduct in the absence of a convincing evidence base,
just as well as in its presence. These rules may be changed if the evidence base or its interpretation improves.

One might say that so-called CAM (complementary and alternative medicine) claims conjectural abilities for
persons performing certain forms of investigation and treatment.%
\footnote{These conjectures are more often than not
disputed by members of the conventional medical profession. What speaks against an outright rejection of
CAM is the large number of of patients who make use of it in some form or another, many of whom are reporting
positive experiences about CAM, a phenomenon which by itself requires an explanation.}

Thus, in order to appreciate the value of EBM one must appreciate that CCBM excludes CAM rather than
that it positively implies a professional's abilities. The EBM fraction of CCBM is probably growing. That is
most likely the case when measured in terms of knowledge. When measuring in terms of numbers of
treatments applied by medical professionals the growth of the relative frequency of EBM is
limited by the fact that successful EBM when applied to a specific case is likely to come to an end whereas
non evidence based CCBM may go on forever without solving the problem. As people live longer, a plausible
consequence of well-organized EBM (almost by definition),  they may develop a greater variety of problems for which no significant evidence base can be generated. In an extreme future development each EBM enabled case is
diagnosed and
treated automatically
by means of computers and robots, and human medical professionals are called in only
when EBM is inapplicable and either non evidence based CCBM must be applied, or CAM is tried. This thought experiment even potentially undermines the objective that medical professionals should increasingly
apply EBM methods. If the difference between non evidence based CCBM and CAM is blurred, which would not
be unreasonable on the long run, the future of human medical professionals may even be mainly, if not exclusively,
outside EBM.

\subsection{Personal and aggregate based competence}
Competence can reside in an individual person or in some aggregate. It may reside
 in a group of named persons, in an organization which due to size and structure has become independent of its individual employees, in an institution comprising many organizations,  and in a community which is large enough to be independent of its particular members. We assume that group competence, community competence and organizational competence are each resulting from two ingredients: (i) the cumulative competence of group members, community members, and institution employees, and (ii) the availability of an accessible information base for the group, community, or institution.%
\footnote{Organizational competence understood as core competence need not be derived in a bottom up way from what employees can do alone and in teams. In \cite{Tampoe1994} a top-down approach for detecting such competences is preferred. It is even claimed in \cite{Tampoe1994} that staff need not be aware of a company's core competences, so
that mining those becomes a non-trivial task which needs to be done in a systematic fashion.}

Below we will focus on individual and personal competences. It is  assumed that these are also a key component of competences  higher levels of aggregation (group, organization, institution, and community). Only in the presence of group members with individual competence can a group show group competence and the building and usage of an adequate and effective information base is also preconditioned on the existence of such persons. The question how to design and develop a useful information base on sourcing and outsourcing is far from trivial and has not yet been solved.

The term competence has many meanings apart from the dictionary entries that already have been mentioned. In
\cite{RobothamJubb1996} doubts are expressed that the phrase management competence is of any
use given the different approaches to its meaning. It is stated that instead of defining competence some of its forms may be specified such as hard and soft competences or threshold and high performance
competences. One needs a definition of competence if it is to be made measurable. That can vary from context to context. In \cite{Shippmann2000} a liberal approach is taken, and competence can refer to
a large range of packages of abilities, skills, cognitive and emotional capabilities, attitudes and convictions.

In the sequel a lightweight interpretation of competence will be used, which includes topical knowledge as
well as topical experience. Thus following the usage below claiming competence is less demanding
than claiming knowledge, skill, ability, comprehension, overview, performance, understanding,
or insight.

\subsection{Survey of the paper}\label{Survey}
Having posed the questions to be addressed by way of a so-called pragmatic perspective on
competence and its critique above, we proceed with a description of what one might call outsourcing practice
here termed professional action embedded sourcing and outsourcing activities.

In Section \ref{SOC} we deal extensively with competence. We begin with a generic competence classification,
which is then instantiated for sourcing and outsourcing related topics.

We are led to the observation that outsourcing in general and domain specific sourcing
(such as IT sourcing), can and
should be distinguished and that neither of these are subsumed under a service industry competence that transpires from the literature on service science.

Then we list a number of issues that come into play if individual acquisition of competence on sourcing and outsourcing is contemplated, including measurement of competence in the absence of evidence based methods.

In Section \ref{SOA} a distinction is made between (i) competence, which in the case of sourcing and
outsourcing is labeled as  community confirmed competence (which may both be evidence based and non evidence based) , (ii) ability which must be evidence based, and so-called conjectural ability. Several classifications of ability
are given, and comments on the acquisition of abilities are provided. 

In section \ref{Options} the question is considered what forms of theory of outsourcing may exist.

In section \ref{ReqToS}  requirements will be formulated on a theory of outsourcing, (yet to be developed), under the assumption that awareness of an
outsourcing theory (compliant with these requirements) contributes in a well-understood way to a person's outsourcing competence profile.

Finally Section \ref{Conclusions} contains concluding remarks.

\subsection{Are we begging the question?}
The set-up of this paper makes it vulnerable to the following objection. If outsourcing competence is assumed as a criterion for defining the paper's audience, and if from that notion further requirements are derived on how to design a theory of outsourcing that enhances outsourcing competence: how to prevent
that the theory subsequently leads to a different definition of outsourcing thus rendering the original audience description as well as the analysis of what that audience needs deprecated.

Of course this problem can be avoided by working out the details of an outsourcing theory first, and
only thereafter presenting the story, but doing that is not our plan, because we intend to work in a systematic
top-down fashion, while leaving open on purpose
where outsourcing theory development is actually going to bring us.

The objection may be rephrased as follows, (i) what can be inferred if some transformation that was thought of as outsourcing fails to comply the criteria that the theory demands once having been
developed in significant detail, and (ii) what  can be inferred if some activities
are considered outsourcing after all that don't match the description on which the original indication of the intended audience was based. Issue (ii) has the simpler answer: if the relevant audience turns out to be
larger than originally thought some additional presentation for that additional audience can be considered.

The first issue is more problematic and in addition its occurrence is more plausible. Indeed we will
classify some activities under IT-sourcing which might be classified as IT-outsourcing by other observers of the outsourcing industry. So the theory that we are looking for is probably  more restrictive than some common interpretations
of the jargon suggest. In that case the theory is not applicable to certain cases in spite of the fact that those
cases are often considered to represent  instances of outsourcing. This limitation is acceptable if a person
$P$ who intends to apply the theory is informed about it.

Besides the chicken and egg problem just mentioned that comes about from specifying an
audience for an outsourcing theory before providing its details, a second matter may give rise to
objections.
Outsourcing is not subsumed under sourcing  but making a firm distinction between sourcing
and outsourcing is a conceptual challenge. Sourcing will be understood from the perspective of a
unit $U$ as taking care of the availability to $U$
of adequate sourcements in order to satisfy $U$'s functional needs.

Sourcement transformation is the subject of transforming sourcements, including the design and
putting into effect of transitions implementing such transformations.

Outsourcing is the most well-known sourcement transformation. Apart from representing a specific sourcement transformation type it often stands for the entire range of sourcement transformations combined with a fundamental awareness of unit mission policies. So in advance of a more detailed theoretical treatment, outsourcing (assuming a context where it also stands for insourcing, backsourcing, and
follow-up sourcing) may be understood as ``mission based sourcement transformation'', or when mission matters less as ``mission aware sourcement transformation''.

\section{Professional action embedded sourcing and outsourcing activities}\label{PAESOA}
The notion of a person $P$'s $X$ competence, understood as the quality of $P$ that $X$ type tasks are
usually performed adequately by $P$, is somehow circular.%
\footnote{Even if one accepts the implication from $P$'s (high) probability to succeed on $X$ tasks to $P$'s
$X$ competence (which we will not do), then this one way implication falls short of defining competence.}
To see this circularity  it is helpful to consider incompetence,
understood as the absence of competence, rather than competence.
If $X$ stands for piloting a glider plane then
claiming $P$'s incompetence  for $X$ (in the absence of an outright security incident during an actual flight piloted by $P$) often requires the availability of a more rudimentary $X$ competence, say $X^{\prime}$,
that is $X$ subsumes $X^{\prime}$, which enables an observer to
create a testing context in which $P$ can fail on performing an $X^{\prime}$ task. The circularity appears through the incentive to find $X^{\prime}$ in order to clarify what incompetence may amount to in the particular case at hand.

We speak of professional action embedded activities%
\footnote{One may consider to replace ``professional action embedded'' by ``professional'' in the context
at hand. That only seemingly leads to be a simplification, however. In this paper professional activity is supposed
not to imply competence, at least not by necessity, though often it will. Indeed a professional
in some domain can be incompetent, which state of affairs may be undesirable but  is not incoherent. It may be an objective to manage a profession $X$ in such a way that all $X$ professionals are $X$ competent. It may be an objective of $X$ professionals as a group to organize themselves and to develop an agreement about what $X$ competence amounts to and to enforce subsequently that in due time all $X$ professionals are certified as to comply with the qualification of having provably acquired that competence. Of course the specification of $X$ competence may be in need of maintenance which can be done under the responsibility of the $X$ professional's organization.
After some evolution of the domain $X$ and its profession,  the very concept of $X$ competence itself,
as laid down at some stage, may enter the definition of an $X$ professional thus disabling the decoupling of these matters. In that case clarification may be obtained by working out a progressively demanding sequence of competence specifications (qualifications) in parallel with a progressively restrictive sequence of descriptions of the profession at hand. These complexities  are avoided  by making use of the phrase ``professional action embedded''.}
in order to indicate that (i) tasks are performed with full concentration and on the basis of adequate preparation,  and (ii) that at least in principle some external agent must be satisfied with the outcome of the activities, and (iii) that the result of the activity takes priority over its learning outcome on the persons involved as well as on their appreciation of the time spent on the matter.

Thus an actor's professional action embedded activities are: (a) not primarily motivated by the learning outcome for the actor, (b) not motivated by a recreational  value for the actor, and (c) have no health or well-being target for the actor except generating normal remuneration. For professional action embedded activity it is (d) not guaranteed that the actor is aware of or applies
best practices in any form, and (e) it is also not implied that  the actor is specifically certified or qualified. Further (f)
no external performance measurement or assessment is required for professional action embedded activity.

We cannot analyze outsourcing competence by first explaining outsourcing tasks and roles and then explaining
what  competence means in these particular cases as if competence does not enter the description of these
roles and tasks to begin with. The solution for this particular circularity lies in abstraction. Providing an abstract
indication of the activities within the target domain at hand, an indication that suffices for an originally
uninitiated person to appreciate may be possible for persons not equipped
with any preliminary competence for the target domain. Some examples may be useful: (i) vertical rock climbing in the Dolomites as an activity. This activity can be explained by means of some text with pictures without
requiring or transferring any competence for it, (ii) driving a motorcycle in an EU country, (iii) outside high altitude
tightrope walking without fall protection and without a balance bar.

Now we will provide an outline of the activities on which this paper will focus. This outline has no pretense of providing complete and adequate definitions, important aspects are missed out. But it provides an abstraction that serves as a starting point.%
\footnote{This is a notorious difficulty regarding the use of non-mathematical definitions.
If we write that a car is a means of
transportation that assertion cannot serve as a definition of a car, because it is insufficiently specific. Nevertheless it is a valid assertion about cars for those who know what a car is already. But if we have not yet encountered any cars how to get started? We will think in terms of an author (or speaker) who knows (or claims to know)
 what a car is and who intends to communicate that knowledge in a structured and top-down fashion, introducing additional detail only when needed. Introducing a car as a means of transportation can serve as a kind of preliminary definition sufficient for initiating a discussion of cars. It may be considered a definition with a low resolution. Perhaps one may speak of a low resolution predefinition, which can be refined during the course of a discussion by predefinitions with a higher resolution until perhaps a ``true'' definition is obtained. Using software engineering terminology one might prefer to think in terms of a specification in some stage of refinement. But in software engineering one specifies a representative of a known class (of systems) by means of its properties. What gets lost when talking of specifications is the intentional aspect. The assertion that
a car is a means of transportation can make sense to an audience of persons who have never seen anything like a car before, and who know that they are probably going to be introduced to a completely novel idea. The transmission of a low resolution predefinition of a car is only a first stage in a process which is supposed to end in a state where the members of the audience have a workable mental picture of a car available, even if no cars yet exist. A (concept forming, or  following the terminology of \cite{BV2010}, an imaginative) definition
then is an abstraction (or better, a description of an abstraction) extracted from  a range of more detailed (higher resolution) mental pictures, preferably collected from different persons. Deciding upon a definition is a matter of individual or of group decision making because
 there is no intrinsic quality of ``being a definition'', which can be checked, though some design rules can be applied. For instance the removal of all intentional aspects adds to the quality of a definition. Predefinitions, however,
 may be stated in intentional terms.}
The terminology of units and sourcements is taken from \cite{BDV2011a}.

\subsection{Outsourcing and insourcing}
Outsourcing is about realizing business transformations where a sourcement is moved outside a unit (the outsourcing unit) while still performing a significant part of its original role (the production of a service)
towards the original unit, for some period to come. Outsourcing activities appear in many different roles from decision making to transition management and from business case analysis to contact design. The group of persons working on an outsourcing task must be fully aware of the impact this has on the outsourcing unit's mission, an adverse impact is inconstistent with the very concept of outsourcing.

Outsourcing takes place in many different domains. Domain specific outsourcing such as IT-outsourcing is specific for a particular domain.

Insourcing is complementary to outsourcing: if a unit incorporates a sourcement in the course of an outsourcing, that unit is said to be insourcing the sourcement.  It is likely that an insourcing organization has a mission with a focus on specific domains, for instance: IT services, IT system development, bookkeeping and financial services,
catering and serving specific foods and drinks, dedicated research and development, managing IPR portfolios.
It is plausible that insourcing is done compatible with the insourcer's  mission but that is not imperative. Insourcing
activities comprises all tasks performed concerning a sourcement transferral on behalf of the insourcing unit.

There is an asymmetry between outsourcing organizations and insourcing organizations. Insourcers may primarily view their own business as service provision with insourcing as a second best option, necessary
to offer when a potential customer is only willing to accept particular services if some of its sources are insourced
by the prospective service provider, thereby reducing the cost for the outsourcer, but at the same time significantly
increasing the complexity for the service provider. No outsourcer, however seems to be in the business of outsourcing as such.%
\footnote{This may be disputed: for a university it may be a useful strategy to develop activities (sourcements) internally to the level that these can be successfully  outsourced. One may think of timetabling processes, research contract management processes, educational online publishing, and highly specialized international staff recruitment. This
ambition to outsource a sourcement for which a unit has been the primary customer is quite different from the conventional breeding of a spin-off company within the institution because the institution constitutes the first and pioneering client for the new sourcements service. Outsourcing of sourcements which are vital (that is mission critical, though not mission defining)
to a unit  may initially take place towards some kind of holding remotely
controlled by the outsourcing institution intending to sell it with profit once new external clients have been found.
Seen from  this perspective it becomes plausible to invest much more effort and funding in sourcements hosting
mission critical competences that are not non-mission defining. This fits well with the explanation of the potential of outsourcing at large in \cite{BDV2011a}.}
Outsourcing is hardly conceivable as a unit's mission itself, while thematic service provision, systematically expanded by successive insourcing operations in order to gain market share, may very well be a unit's mission. If one modifies outsourcing into service consumption the situation does not change: service consumption (whether or not restricted to a specific domain) cannot constitute the mission of a unit. We find that outsourcing and insourcing have a quite different relation to the mission of the unit which performs the sourcement transformation.%
\footnote{So, whereas professional insourcing might be viewed as an add on feature of the service industry,
if not as a kind of service itself, outsourcing has a much more independent status.}

\subsection{Backsourcing and follow-up outsourcing}
Essential for outsourcing is the temporary nature of the business relation. Backsourcing is a partial inverse of outsourcing which may be encountered after the task allocation resulting from an outsourcing has come to an end, for instance because of contract expiration. Instead of backsourcing, besides task termination, follow-up outsourcing is an option. That involves finding a new unit and moving towards a situation as if the outsourcing had been performed towards that new unit.

\subsection{Sourcing}
Sourcing  involves all activities intended to make appropriate sources available for a
unit and to maintain their good order. The difference between sourcing and procurement is not always appreciated, but we hold that procurement may be considered a part of sourcing which has some bias towards the initial phase of sourcing, that is making sources available, while sourcing expressively concerns the question which sources a unit is best advised to avail itself of. Sourcing professionals must think in terms of a sourcing strategy for a specific domain, for instance IT sourcing. For a complex organization sourcing activities require up-to-date knowledge of developments in the relevant service industries as well as spotting technological progress which may indicate  plausible insourcing transitions or which enables backsourcing  (though in a different form) business processes that had previously been outsourced because of their dependence on complex technologies.

We hold that there is no general or generic sourcing profession because that is too general.
There are many thematic sourcing competences, however. In contrast there is a generic outsourcing profession,
which includes backsourcing and follow-up outsourcing activities as well. Probably there exists no general insourcing profession but there are domain specific insourcing areas with a sufficient size to allow professional development.

\subsection{Domain specific sourcing and outsourcing}
Given a  domain, say IT, one may reconsider
 the above survey of activities and contemplate the following dedicated fields of specific activity.
\begin{description}
\item{\em IT outsourcing.} This has been an important subject, often referred to as ITO, but it is gradually overtaken by  IT sourcing, because many organizations have already outsourced a large fraction of their IT sourcements. Further the aspect of outsourcing gradually diminishes in favor of a less constrained process of making use of IT services.
\item{\em IT insourcing.} This has been an important activity enabling the rapid growth of some international service providers. However, cost effective service provision without any promise to insource any of the outsourcer's sourcements is becoming more prominent.
\item{\em IT sourcing.} This is a very important activity, rapidly developing into a profession in support of general IT management, needed by many organizations.
\item{\em IT sourcing consultancy.} This is an equally important activity, rapidly developing into a profession in support of general IT management, needed by many organizations. The point is that a specialized IT sourcing consultant can become more experienced in IT sourcing than a specialist operating within company
bounds, especially if the company is either relatively small or if it is mainly focused on objectives outside IT. What adds to the plausibility of IT sourcing consulting is the rapid speed of technology change in IT. Even taking notice of this dynamics may be unfeasible for a small or medium-sized organization. Working together may be blocked
by legislation on fair competition.
\item{\em IT outsourcing consultancy.} This activity is widespread because outsourcing may be incidental for the outsourcer and may not be part of its normal business model so that it cannot entirely
be handled by a company's own crew. IT insourcing consultancy is a less plausible activity because insourcing is likely to be  part of the insourcer's business model.
\end{description}

\subsection{Sourcing and outsourcing versus servicing}
The viewpoint that sourcing is the modern term for what used to be outsourcing, and that a sourcing competence must be acquired and maintained instead of an outsourcing competence, seems to be based on a bias towards thematic competences. Indeed, IT sourcing seems to have become more prominent as a descriptor of roles and tasks than IT outsourcing, but from that observation it cannot be concluded that (general and domain independent) sourcing has become more prominent than (general) outsourcing. Based on these considerations the following grouping of activities is plausible.
\begin{description}
\item{\em general outsourcing.} A key activity for many large organizations.
\item{\em domain specific sourcing.} IT sourcing is a paradigmatic example of a domain
specific sourcing area. Financial and administrative process sourcing is plausible too. Indeed various domain
specific sourcing competences (perhaps not always explicitly labeled as such)
are essential for many organizations and for many consulting units.
\item{\em domain specific service design, engineering, and deployment.} This is the subject of service science.%
\footnote{It might be termed servicing in order to make the terminology more coherent.}
From this perspective insourcing is merely a strategy for service providers who are not sufficiently in command of their market. Service-dominant logic (see \cite{VargoLusch2004}) provides the professional language of this world which considers outsourcing and insourcing to be of a secondary importance only.
\end{description}

Of course one might contemplate a general sourcing professional, but achieving this role is an unrealistic expectation for any individual. Based on these considerations the term outsourcing still deserves a special place in (or next to) the sourcing landscape. IT being very prominent in the family of domains admitting substantial domain sourcing competences it is clear that IT sourcing plays a special role if only by being the paradigmatic example of thematic sourcing, but it does not subsume (general) outsourcing, neither does it subsume IT outsourcing, in both cases because it operates at a different (that is lower) level of abstraction.

\section{S\&O competence}\label{SOC}
In order to develop clarity about sourcing competence and outsourcing competence a generic method
for naming competences will be proposed. The leading idea is that $X$ competence consists of experience
with professional action embedded $X$ activities in the light of Section \ref{PAESOA}.%
\footnote{This idea come from the assumption that having seen successes and failures during the
 period of obtaining experience a person $P$ can us analogies (see \cite{Waller2001}) to preceding cases,
 and inferences from theoretical classifications and definitions (see \cite{WaltonMacagno2010}) in order
 to explain new cases. Now in \cite{Trout2002} one finds a warning concerning the assumption that mere experience
 will suffice in the absence of methodical collection and aggregation of the outcomes of that experience.
 Hindsight bias  and overconfidence bias will be a treat for $P$ if no methodical support is used. Hindsight bias
 may lead someone to overestimate the explanatory value of having selected a particular course of action for the good or bad outcome of a preceding case, thus overestimating the plausibility of repeating or avoiding that or a similar course of action in a current case. Overconfidence bias may be unhelpful because it may lead to someone being sure of one's decisions on too marginal grounds. Competence acquired by experience is not an effective remedy against overconfidence. That bias needs a much more sophisticated remedy.}
Several other notions must be contemplated simultaneously at this stage: competence, ability, capability, qualification, certification, registration, possession of a diploma, possession of a certificate. Ability will
be discussed in the next Section. Capability is a mix of ability and competence. A qualification $Q$ is a
degree of competence, a $Q$ qualified person has been examined to meet the requirements of the
qualification $Q$. Certification consists of deciding upon a package of qualifications and upon methods for
validating that a person meets the qualifications involved. Certification, is more than the mere
awarding of a certificate, it involves the intention that the certificate implies some form of guarantee that a
certified person can perform in a competent way in some domain. A diploma is a certificate awarded by an
educational institution, in comparison with a certificate issued by a professional society or it is awarded on the
basis of a specialized training scheme. A diploma has more focus on indicating a competence level of the receiving
person than on that level being sufficient for adequately performing certain tasks.

\subsection{Generic competence scheme}
For a domain $X$ a number of related competences, or classes of competences,
can be distinguished. A structured overview of these will be
called a generic competence scheme. The scheme is structured by the subsumption relation.
One competence may subsume another competence.%
\footnote{In object oriented terms: if competence $A$ subsumes competence $B$, the class of $A$
competent agents inherits (and thus extends, in the sense of class extensions) the class of $B$
competent agents. In set theoretic terms if competence $A$ subsumes competence $B$, the set of $A$ competent
individuals includes that of the $B$ competent individuals.  Inheritance can be multiple,
that is a competence can subsume several other competences at the same time.}

Thus with a topic $X$ comes
a range of $X$ related competences. Competence appears in two meanings everywhere: depending on how it occurs in a sentence $X$ competence is either (i) a class of competences (viewed as a  single competence in its entirety) constituting the sum of
someone's (or a group's etc.) $X$ competences, or (ii) it refers the members of that class, that is one of the
the diverse $X$ competences that together make-up $X$ competence (in the earlier sense). A member of the class will be referred to as a micro competence. Thus: (i) $X$ competence is the sum of the  micro $X$ competences,
and (ii) an $X$ competence is a micro $X$ competence.

For each micro $X$ competence one may
expect to assign a degree of sub-specialization. If that degree is very low the notion is uniquely related to $X$ in its entirety. If that degree is high $X$ competence refers to a sub-speciality rather than to $X$ as a whole.%
\footnote{The term sub-specialization has been chosen because $X$ is already a specialization.  When writing that backwards parking is an important car driving competence, that occurrence of competence has a high
degree of sub-specialization. But when writing that taxi drivers have above average car driving competence the occurrence of competence has a low degree of sub-specialization.}
It is understood that a person's $X$ competence (in the sense of a low sub-specialization) is a weighted sum
of a persons (high degree sub-specialized) $X$ competences.%
\footnote{Determining these weights may be a matter of community based decision making itself
based on more or less shared views of what is important, central, or effective. It may involve specialized
theories on skill acquisition and measurement.}
The following listing is fairly inclusive but need not be exhaustive.
\begin{description}
\item{\em $X$ framework competence.} Preliminary and approximate knowledge of key concepts of $X$
which together constitute a framework from which to approach $X$ as a topic.
\item{\em Theoretical $X$ literacy competence.} Extends  $X$ framework
competence with awareness of one or more approaches to a theory of $X$.
\item{\em  Empirical $X$ literacy competence.} The competence that arises from substantial awareness of
empirical research on $X$.
\item{\em  Empirical $X$ research competence.} The competence that arises from working as a researcher
performing empirical research on $X$.
\item{\em  Theoretical $X$ research competence.} The competence that arises from working as a researcher
performing theoretical research on $X$.
\item{\em Domain $Y$ specific $X$ competence.} The competence that comes about from participation in
$X$ projects or activities that have a dominant focus within either subdomain $Y$ of $X$
or $Y$ directed usage of $X$ activities. This competence class may
include a propensity to follow some specific (accepted) rules of conduct, to apply some specific (accepted) methods,
and not to apply some specific (rejected) methods.
\item{\em Theory informed domain $Y$ specific $X$ competence.} An extension of domain $Y$ specific $X$ competence for some domain $Y$ combined with theoretical $X$ competence and with a
specialization of the theory at hand to domain specific concepts for $Y$.
\item{\em $X$ competence.} Domain $Y$ specific $X$ competence in at least one domain $Y$.
\item{\em Evidence based $X$ competence.} $X$ competence consisting of experience
with evidence based practices concerning $X$.%
\footnote{Evidence based $X$ competence as a phrase indicates a special case of $X$ competence. In other words speaking of evidence based $X$ competence presupposes the existence of an $X$-competence
which need not be evidence based. This is a weaker form of competence which can evolve earlier in time. Determining what that weaker competence may amount to is a task that may vary from topic to topic. Between $X$ competence and evidence based $X$ competence one may imagine a continuum of intermediate competences,
for which no naming scheme is provided.}
\item{\em Theory informed evidence based $X$ competence.} An evidence based $X$ competence augmented with
 theoretical and empirical literacy competences sufficiently developed to gain a reliable understanding of the
 evidence base of the $X$ competence at hand.
\item{\em General $X$  competence.} The simultaneous availability of two or more domain
specific $X$ competences.
\item{\em Theory informed general $X$ competence.} A theory informed combination of
three or more theory informed domain specific $X$ competences for different domains.
\item{\em $X$ consultancy competence.} Extends empirical $X$ literacy competence with $X$ consulting experience.%
\footnote{In some domains it is reasonable to assume that $X$ consultancy competence
subsumes $X$ competence.}
\item{\em Community confirmed $X$ competence.} An $X$ competence which in addition
has acquired community confirmation.

\item{\em Community confirmed domain $Y$ specific  $X$ competence.} An domain $Y$ specific $X$ competence which has acquired community confirmation.
\end{description}

\subsubsection{Competence subsumption relations}
We notice the following subsumption relations that will hold for all domains $X$:
\begin{itemize}
\item Every other $X$ competence subsumes $X$ framework competence.
\item Theoretical $X$ research competence subsumes theoretical $X$  literacy competence.
\item Empirical $X$ research competence subsumes $Y$ specific empirical $X$ literacy
competence for one domain $Y$ at least.
\item General $X$ competence subsumes domain $Y$ specific $X$ competence for at least
two different domains $Y$, provided more than one subdomain is distinguished within the domain $X$. Otherwise $X$ competence and general $X$ competence coincide.%
\footnote{Obviously taking the bound two is an arbitrary decision.
Perhaps this bound must be increased dependent on $X$.
There is some lack of precision here: whoever is generally $X$ competent is domain $Y$ specific
$X$ competent for at least two different domains $Y$ provided these can be found within domain $X$.}
\item Theory informed general $X$ competence subsumes domain $Y$ specific
theoretical $X$ competence  for at least two different domains $Y$, as well as theoretical $X$ competence.
\end{itemize}

\subsubsection{About incompetence}
We will speak of $P$'s $X$ incompetence if some of the micro $X$ competences that make up $X$ competence are
missing for $P$. This definition applies only if $X$ competence is understood as a combination of micro competences
for a fixed set of subdomains. This pattern of explanation of incompetence is mainly
applicable in  $X$ related competences just
mentioned except the ones involving community confirmation.

For $P$ to feature  community confirmed $X$ incompetence, there must be some micro $X$ competence on which
$P$ demonstrably fails and that demonstration is community confirmed as a demonstration. This failure can be
a lack of specific experience, the failure to apply a specific (accepted) method, or the application of a method of which the rejection is community confirmed.

\subsubsection{Competence profile}
A person $P$'s confirmed $X$ community competence profile consists of a survey of $P$'s community confirmed
$X$ competences. This is independent of the way in which $X$ community confirmation is organized in a
particular community which may range from very informal to very formal. The community competence profile
includes both evidence based elements and non-evidence based elements.

Two phenomena are not included in a community competence profile while their existence must be acknowledged at his stage already. Firstly and most importantly, evidence based competences (below also referred to as objective abilities
and also simply as abilities) that
have not succeeded to achieve community endorsement are not included in the community confirmed
competence profile. Secondly, special interest groups that do not reflect even a significant minority
of an $X$ community may maintain a package of abilities, often derived from a theory which' adherence is
constitutive of the interest group. Insofar as those abilities are lacking an evidence base their status is necessarily ``conjectural'' only. Conjectural abilities may be community confirmed%
\footnote{The ability of the (personnel of) the penitentiary system
to reduce crime by means of strict penalties may be an example of a community confirmed ability
that may lack a sufficient evidence base.}
or may lack community confirmation.
If an evidence base exists for an ability the status is ``objective''.

Because we understand competence by default (that is when occurring without further adjectives) as community competence both non-community confirmed but
evidence based ability (in brief non-community confirmed ability) and non-community confirmed conjectural ability
(commonly promoted by a special interest group lacking community  wide visibilty) fall outside the community
competence profile. These ``abilities'' will be included in the so-called competence profile, however. A competence profile results from including these two forms of ability as well.

The resulting jargon is perhaps grammatically not fully satisfactory because (i) it implies that a competence profile includes elements not understood as competences (because evidence based but not community confirmed abilities fall outside competences on the basis of interpreting competence primarily as community confirmed competence),  and (ii) non evidence based
conjectural abilities (if at all accepted as a category close to abilities) are unlikely to acquire community confirmation.

\subsection{Specialization to sourcing and outsourcing}
Having some generalities on the concept of competence at hand the next step is to specialize these generalities to the subject of outsourcing, which is our main objective, and as a byproduct to the subject of sourcing.

\subsubsection{What is specific for outsourcing concerning its competence?}
Focusing on outsourcing, one may ask what aspects of that subject may create a setting in which the
concept of outsourcing competence requires an analysis that may differ from competence
analysis for other some areas of human activity. Here are some aspects.
\begin{itemize}
\item Scientific research concerning the validity of community competences on outsourcing takes many years.
A sourcement may easily last for ten years and investigating its full life-cycle may take longer. Even if the community is able to agree on the hypothesis that under certain conditions some specific course of action is
advantageous, promoting the knowledge about that rule of behavior to the status of evidence based knowledge
(assuming that that will eventually happen) may
take longer than the duration of a single person's entire career. Scientific progress is slow in this subject, at least when measured in terms of its construction of an evidence based catalogue of competences.
\item The overall business context is changing so fast that it becomes doubtful that an evidence base can be
established for many practices which seem to be good. This is almost paradoxical. The very speed of social
change and technology development may imply that an evidence base cannot be properly obtained when it still matters.
\item Because economic needs coincide for many units synchronously, outsourcing takes place in trends and fashions. A unit's management cannot escape the need to contemplate outsourcing and  it cannot wait until an evidence base is obtained for a particular theory from which its plausibility may be inferred.
The need to act is felt more strongly than the need
to act in an evidence based fashion.
\end{itemize}

\subsubsection{Instantiation of the generic competence scheme}
Many organizations have procurement departments where outsourcing processes are designed and performed on a regular basis. We made an attempt to capture this part of human practice as professional action embedded
sourcing and outsourcing activities. It will now be assumed that this practice is a known entity from which
further notions like outsourcing competence may be derived irrespective of the circularity which that position may
potentially be infected with.%
\footnote{So it is meaningful to say that (i) $X$ practitioners must reflect upon their views concerning
$X$ competence and that, (ii) $X$ practitioners must preferably be $X$ competent, and that (iii) evidence
based $X$ competences work in $X$ practice, but it is not meaningful
(in our use of the terminology) to state or suggest that (iv) the $X$ competent persons as a community
must see to it that an $X$ practice comes into existence, or to even to say that (v) $X$ competent persons must
see to it that they get involved in $X$ practice.}

In terms of competences and professionalism outsourcing is a coherent subject and it is very conceivable that staff having a wide experience in managing outsourcing processes have not been involved in any insourcing processes. It seems to be the case that outsourcing competence, insourcing competence and sourcing competence are quite different and unlikely to be found as competences of the same individual. So it is plausible to consider
instances of the generic competence scheme with sourcing and with outsourcing independently.

For $X$ we will make use of the following substitutions: sourcing, insourcing, outsourcing, backsourcing,
and follow-up outsourcing. All of the resulting competences require preliminary descriptions, for instance
sourcing framework competence involves awareness of the basic notions of (i) units, themes, and processes,
and (ii) service usage and provision.

Now it becomes possible to enrich the relation between sourcing and outsourcing
by means of the following additional subsumption assumptions.
\begin{itemize}
\item Outsourcing framework competence, insourcing framework competence, backsourcing framework competence and follow-up outsourcing framework competence are all the same.
Outsourcing framework
competence is used as a representative for each of these.%
\footnote{An alternative is to speak of sourcing transformation framework competence.}
\item Theoretical outsourcing literacy competence, theoretical insourcing literacy competence, theoretical backsourcing literacy competence and theoretical follow-up outsourcing literacy competence are all the same.%
\footnote{But outsourcing competence, insourcing competence,  backsourcing competence, and
follow-up outsourcing are pairwise incomparable (neither one subsumed in the other), though not pairwise disjoint.}
Theoretical outsourcing literacy
competence is used as a representative for these competences.%
\footnote{An alternative is to speak of theoretical sourcing transformation literacy competence.}
\item Sourcing framework competence subsumes outsourcing framework competence.
\item Theoretical sourcing literacy competence subsumes theoretical outsourcing literacy competence.
\item Domain $Y$ specific sourcing competence does not subsume outsourcing framework
competence.%
\footnote{For any domain $Y$ except the somewhat artificial domain of ``maintaining knowledge about either  general or one or more domain specific outsourcing competences''.}
\end{itemize}

These rules imply that all of the generated competences subsume outsourcing framework competence, while still sourcing competence and outsourcing competence differ.

\subsubsection{Outsourcing framework competence}
Outsourcing framework competence involves knowledge of the content of Section \ref{PAESOA} plus some
awareness of literature on outsourcing plus
some experience with the outcome of outsourcing processes from the perspective of the outsourcing unit, the insourcing unit, or a consultancy role.

Sourcing framework competence extends outsourcing framework competence with awareness of service science
and S-D logic  (service-dominant logic) as proposed in \cite{VargoLusch2004}.

\subsection{Outsourcing theory audience description}
If a competence profile $C$ has been named and a way to decide about which persons are $C$
competent has been specified, the very notion $C$ can be used as a specification of a group $G=G_C$
of persons. Those persons who are in possession of competence $C$ are included in group $G_C.$

The audience specification for an outsourcing theory (yet to be designed) is simply as follows:
those persons who have already
acquired an outsourcing framework competence, that is $G_{OFC}$. Members of $G_{OFC}$
are the target audience in the sense that they should be able to understand and appreciate the theoretical work.  The ambition for developing an outsourcing theory
lies in providing the step from (general or domain specific) outsourcing competence to a theory informed extension of (general or domain specific) outsourcing competence.%
\footnote{The expectation that a
progression to a theory informed stage of outsourcing competence is rewarding from the viewpoint
of a less theory informed stage of outsourcing competence is understood as a design intention, or a requirement concerning the design rationale for an outsourcing theory.
Making sure that this reward is indeed achieved cannot be considered to constitute
a part of theoretical work on outsourcing, as
it essentially involves proving that there is an evidence base for the productivity increase resulting from the move
towards a theory informed competence level. Proving the existence of an evidence base requires
empirical work to be done, and indeed it presupposes the existence of a volume of independent empirical
studies that allows for carrying out convincing meta-studies.}

Whereas an audience description for an outsourcing theory can be formulated at this point, a specification of why
that audience might be well advised to take notice of an outsourcing theory cannot be given at this point. For doing so we will need an extensive preparatory analysis of the notion of ability and its many ramifications.

\subsection{S\&O  competence acquisition}
All competences in any of the topics mentioned in Section \ref{PAESOA} concerning sourcing or outsourcing
will now be collected under the label S\&O (Sourcing and Outsourcing) competence. When needed
the observations below can be specialized to subjects within S\&O such as outsourcing, backsourcing,
and follow-up outsourcing.

Not so much can be said about the institutional acquisition of S\&O competence,
at least not in terms of established educational structures. To the best of our knowledge
there are no academic programs leading to either a BSc or an MSc with a primary focus on S\&O.%
\footnote{In The Netherlands courses on S\&O have been delivered through an initiative of PON (Platform
Outsourcing Nederland)
and under the responsibility of various institutions of Higher Education, in particular the
HvA (Hogeschool van Amsterdam). Further in many BSc and MSc curricula in either business and management, economics, or IT, the option exists to perform a project in S\&O.}

We will now list a number of observations, concerning S\&O competence acquisition, in some cases unfortunately  qualifying as no more than opinions.

\subsubsection{S\&O community confirmed objective abilities}
Objective and subjective abilities will be discussed in detail in the next Section. It will turn out that instances of both
subjective and objective abilities can (but need not) be
included in community confirmed competences. We are unaware of any evidence base concerning S\&O with sufficient stability that objective abilities specific for S\&O can be founded on it. For that reason there are no community confirmed objective abilities concerning S\&O either.

When hiring an additional workforce for an S\&O oriented project, rather than finding workers showing validated (or potentially validated) S\&O competence one hopes to find workers who are likely to be successful
when contributing to forthcoming S\&O projects.  This, however, implies a search for objective ability which we
consider to be currently unfeasible in the area of S\&O. That argument puts competences clearly in place as a
criterion for task oriented personnel selection.

\subsubsection{S\&O community confirmed conjectural abilities} Community confirmed conjectural abilities will also be discussed in the next Section. These are included in the community confirmed competences.

Although there are clear intuitive grounds to appreciate a person who has significant (practical)
S\&O competences when some job involving sourcing or outsourcing needs to be performed there is no evidence
available that such persons perform better than others of equal general (that is non S\&O)
level of experience and competence. All community confirmed S\&O competences seem to be so-called
community confirmed conjectural abilities only.
\subsubsection{S\&O community confirmed competences}
The major source of competence in S\&O comes from practical experience. Such experience is linked to
the professional action embedded activities on S\&O listed in Section \ref{PAESOA}.
We make the following observations on that matter.
\begin{enumerate}
\item The essential notion is that of an S\&O process. An S\&O process can be distinguished if some form
of life-cycle is known for it. In fact for different kinds of such processes different life-cycles must be distinguished.

Given a life-cycle model for some specific kind of S\&O process it can be determined where such processes
are active and for which duration and in what phase.%
\footnote{We will not commit ourselves to any particular sourcing transformation or sourcing activity and
neither to a specific life-cycle model for it, but we mention some phases that apply in the case of
outsourcing: sourcing problem statement, sourcing solution architecture, specification of
transformation, request for proposals, vendor selection, contracting, transition, contract management.}

\item A second essential notion is that of a role in which someone may be active during one or more
phases of an S\&O process. Such roles may for instance be:
project team member (which leads to a variety of well-known roles within a project) for vendor selection,
or demand manager of a sourcement within a unit before outsourcing, or contract manager of a sourcement from the provider side after transition.

\item Equipped with a role classification as well as a phase classification for S\&O processes it is
possible to determine a person's experience with S\&O practice, in terms of the number, duration,
and variety of roles fulfilled   in the context of S\&O processes. In addition becomes  possible,
at least in principle,  to express the volume of experience in both  a quantitative an a qualitative way.

\item S\&O community confirmed competence is present (according to some
formal or informal accreditation issuing  body $B$) in a person if: (i) (s)he has been consciously
involved in S\&O processes and if that involvement exceeds some threshold (as determined
by $B$), and (ii) this degree of experience has been maintained in at least a minimal fashion
(as judged by $B$). Defining this initial threshold as well as the need for regular S\&O
process experience for formal competence maintenance is a
matter which must be addressed by  profession representing  organizations (here represented by $B$).

\item By default S\&O competence will be understood as S\&O community confirmed competence.

\item A competent person thus defined can operate on the basis of some awareness of evidence based on practical experience.%
\footnote{However, evidence of this form has not been validated by any method of validation that has come about from a standardized research process. The latter must in no way be seen as an obstacle for speaking of competence per se. It is, however, a prohibitive obstacle if one intends to speak of evidence based competences (or practices).}

\item
Evidence for a person's possession of general or domain specific S\&O competence, may include other aspects than having played roles in practical business transformation processes. It may transpire in different ways as follows:
\begin{itemize}
\item
By being able to give informative (oral or written) presentations concerning a number of relevant cases,
and by having obtained invitations for doing so.
\item
By having been able to sell one's time as an S\&O consultant to units involved in
sourcement transformations.
\item By having been offered jobs with a focus on S\&O projects.
\end{itemize}

\item
The presence of a person's empirical or theoretical  literacy or research competence (four cases)
concerning S\&O may become visible indirectly:
\begin{itemize}
\item
Publications on the subject of S\&O.
\item
Invitations  for presenting one's knowledge in conferences and workshops, made with reference to
previous knowledge based and knowledge oriented activities.
\item
Academic degrees of various levels, in combination with having authored thesis works on
the topic of specialization.
\end{itemize}
We hold that in the present stage of development of the field
none of these indications of competence can be used as a substitute for general S\&O
competence.
\end{enumerate}

\section{S\&O ability}\label{SOA}
We will assume the following definition of ability: a person $P$'s ability for activity class $X$ measures (that is:
positively correlates with) the expectation of success that $P$ should be assigned in advance by an omniscient observer, when $P$ is asked to perform on a task of class $X$.%
\footnote{This definition has been made independent of actual
observers and it also takes care of the situation that $P$ is unaware of his/her own ability regarding tasks for activity class $X$. An alternative definition can be found by adapting the definition of an indication in \cite{Thompson1956}:
``An ability is that quality of a person that describes (part of) their performative possibilities''.}

This understanding of ability produces significant obstacles if a non-omniscient observer is asked to
make an assessment of a person's ability. If person $P$ is (known to be) community competent in activity class
$X$ it is a common assumption that this fact will entail $P$ to have
an above average ability concerning activity class $X$. That assumption is a mere hypothesis, it does not follow
from the definitions of community competence and of ability.

\subsection{Competence versus ability in the case of outsourcing}
A compelling reason for being so specific about competence and ability is that in the area of sourcing and
outsourcing we will propose a fairly liberal interpretation of the concept of
competence which need not imply or depend on
any evidence for the presence of subjective or objective ability. This caution is needed because
otherwise relevant competences (for sourcing and outsourcing) can hardly be reliably observed  due to a lacking evidence base. This may simply be a matter of time, but it may take some 50 years or more before an adequate amount of evidence has been collected which can support an improved
definition of competence concerning outsourcing which is closer to the notion of
ability (by definition evidence based) to
which it's acquisition is supposed to give rise.%
\footnote{In \cite{PfefferSutton2006} one finds an explanation of why evidence is so hard to obtain in a management setting.}

Perhaps redundantly we stress that we consider it unhelpful to arrange the concepts at hand, either explicitly or implicitly,
in such a way that only evidence based competence is acknowledged. There are different
ways of obtaining competence on sourcing and outsourcing which must be used in the meantime, given today's importance of the subject.

If a competence is not evidence based we will label it as a community confirmed competence, (or alternatively
a community based competence). Community confirmed competence may be considered a stage preceding evidence based competence. Community confirmed competences may or may not be evidence based, and conversely evidence based competences may or may not be community confirmed. Community confirmation is an
informal notion which may but need not involve some form of certification.
The need to deal with non evidence based community confirmed competence cannot be
removed by adopting a philosophy of science that rejects it. It can be removed, in principle and
preferably, by scientific progress.

Even if one deplores the situation that evidence about the effectiveness of
the application of S\&O community competences is very difficult, if not impossible, to obtain,
 it is nevertheless unrealistic to consider scientific research to provide the fast and necessary way out of
this dilemma. The transition from community based competence to evidence based competence
can take many years and depends on the successful completion of  research projects.%
\footnote{The situation may be illustrated with the well-known example of the medical
profession where evidence based practice is slowly but steadily becoming the norm.
Of course, given the importance of nursing, evidence based nursing is needed as well, but its
development has begun later. Given the importance of hospital management, evidence based
hospital
management is very welcome too, at least in principle. But must one wait with contemplating hospital
management competence until that particular brand of competence has become evidence based, or is
it better to invest in community based hospital management competence in the meantime?
The latter seems to be preferable to leaving hospital management competence entirely unanalyzed.}

\subsection{A classification of abilities}
Assigning ability to a person is more demanding  than assigning competence, because competence can be measured in terms of a persons history and experience and does not involve prediction of operational
performance. In the absence of convincing methods to determine abilities one may deal with ability in a
hypothetical way. We will find that
conjectural ability (alternatively: hypothetically postulated ability) is a meaningful addendum to  community competence. A classification of ability in terms of its status  from an observer's perspective is helpful.
To that end abilities of $P$ for activity class $X$ can be classified as follows:
\begin{description}
\item{\em Objective $X$ ability.} It has been demonstrated that $P$ (or persons with a
similar experience and training to that of $P$) have above average success with tasks of type $X$. All evidence based abilities are classified as objective. We will understand ability as objective ability by default, that is in the
absence of another qualifier such as subjective or conjectural.

\item{\em Subjective $X$ ability.} (Alternatively: intersubjective conjectural ability,
and also community confirmed conjectural ability.)
The fact of $P$'s ability is known to the public (perhaps including $P$, possibly
represented by some key observers)
because a significant majority believes that some package of $X$ competences and skills,
has prepared $P$ to be successful above average for tasks of class $X$.%
\footnote{In \cite{Thompson1956} the need for the concept of subjective ability is clarified in the context of archeology.
However remote that topic may be from outsourcing, getting a clear conception of the concept of competence is at least as difficult for archeology as it is for outsourcing. A subjective ability may be objective at the same time.

In spite of the frequent occurrence of subjective competence in literature on competence it is a somewhat problematic concept due to its tricky dependency on the community involved. If we consider a community of persons none of whom masters any Swahili, someone's competence to communicate in Swahili may be unnoticed by some part of the community. At the same time it may be conjectured by an other part of the community lacking conclusive information to that extent. This should be considered an example of an objective competence which is not subjective, because the presence of the competence can be easily tested. }
Subjective $X$ ability may take a very different form for different persons.%
\footnote{In \cite{Schettkat2010} the natural rate theory is put forward as a theory that generates a conjectural ability
to steer national economies which lacks an evidence base, claimed by politicians as well as national banking functionaries, and seemingly in command of community approval. In that view the main policy receipt following from natural rate theory, namely that labor market reforms constitute the only path to progress (because monetary policy
based incentives produce temporary fluctuations only) qualifies as a subjective ability.}

\item{\em Established $X$ ability.} Established abilities are subjective and objective at the same time.

\item{\em Auto-subjective conjectural $X$ ability}
(Alternatively: self-endorsed conjectural $X$ ability.) On the basis of some specific experience, training,
or theoretical information $P$ hypothesizes its own (above average) ability for tasks of class $X$.

\item{\em Exclusive intersubjective conjectural $X$ ability.} (Alternatively: interest group endorsed conjectural $X$
ability,  or non community confirmed conjectural ability.) $P$'s ability is conjectured, that is  presumed as a hypothesis, by a minority of community members (excluding $P$).
This group acts as a special interest group working in favor of
some specific qualification processes which may not yet have been endorsed by the community (of
type $X$ task workers) at large. This hypothesis is based on the members of that minority having observed $P$'s knowledge, training, experience or theoretical informedness, together with their belief that these observations
justify the belief of $P$'s ability.%
\footnote{Here it is assumed that belief implies a weaker form of certainty than knowledge. That may be disputed.
If one regards belief as undisputed knowledge the phrasing must be changed to the effect that the justification
is inferred, from the mentioned observations, for the conjecture of ability rather than for the belief of it.}

\item{\em Emergent conjectural $X$ ability.}  Some group may contemplate abilities that have not yet reached the
stage that some persons are believed to avail of those. Clearly this contemplation requires that an ability is abstracted from the persons in command of it. Such an abilitiy, if realistic in the sense that in some
foreseeable future that situation may change may be called emergent.%
\footnote{The competence of simultaneously mastering two different natural languages for which no such
person is known yet provides an example. A newly designed educational scheme may also give rise to persons
acquiring emergent conjectural competences.}
\end{description}

\subsubsection{Conjectural ability versus subjective ability}
Unqualified abilities include all forms of qualified abilities.  Now the notion of a conjectural ability
relates to subjective ability and objective ability as follows:
an unqualified ability (that is explicitly ignoring the default rule that an unqualified ability is objective)
 $\alpha$ of  a person $P$ is conjectural if: (i) it is not an  objective ability, and (ii)
 the interest group $G$ (that is $P$ itself in case of 'auto-subjective', the community or its key representatives in case of 'community confirmed' or  some interest group in case of 'exclusive') assigning $\alpha$ to $P$ know that it is not an objective ability, (iii) $\alpha$ is accepted by $G$ as a candidate subjective (or objective) competence,
 and finally $G$  believes that $\alpha$ may be plausibly assigned to $P$.

\subsubsection{Inability}
$X$ inability is the absence of $X$ ability.
Objective $X$ inability comes with a proof of inability. The absence of a subjective ability may or may not
itself be an objective matter. Subjective $X$ inability is the community confirmed assertion
(or conjecture) of $X$ inability. In principle one may speak of conjectural inabilities, but those
cannot be acquired by becoming  aware of a theory. A conjectural $X$ inability is present with $P$ if $P$'s competence profile does not contain $P$.

\subsection{Ability (unqualified) based on awareness of research results}
The distinction between competence and ability is needed for explaining what contribution empirical and or theoretical $X$ research competence may deliver to a person's
community competence profile. This will now be discussed in the particular case where topic
$X$ is outsourcing and for theoretical research competence including theoretical literacy
competence on outsourcing.

An application of this arrangement of concepts is found when contemplating  the question how to justify
a person $P$'s time spent on taking notice of a particular theory $T$ of outsourcing, which has been
advanced by a team $A$ of authors. There are five scenario's, of which the fifth one is most plausible. Formulating these scenarios except the first two requires that a concept of ability can be made use of:
\begin{enumerate}
\item The team $A$ has been able to convince a majority of the audience of $X$ competent persons that
knowledge of $T$ contributes to this competence. In this case awareness of $T$ can be included as an item in the description of $X$ competence.

Here $P$'s
knowledge of $T$ generates a subjective conjectural ability of $P$.

\item The team $A$ has been able to convince a majority of the audience of $X$ competent persons that
knowledge of $T$ contributes to this competence, though this majority is unaware of the content of $T$. This
may happen if $T$ has been published in certified outlets for research on $X$ so that the majority of the audience for $X$ has a justified belief in the quality of the work that led to $T$ without having made an effort to acquire any awareness of $T$. In this case awareness of $T$ can be included as an optional item in the
description of $X$ competence. The certified publication of $T$ justifies the assumption that taking notice of it
will not have detrimental effects.

Here $P$'s
knowledge of $T$ generates a authority based subjective conjectural ability of $P$.

\item In some cases it can be demonstrated that mastering theory $T$ adds to the
ability of $P$ for class $X$ tasks. This is an exclusive state of affairs, however, and in the subject of outsourcing it
may be quite far away in the future that such demonstrations will occur.

Here $P$'s knowledge of $T$ generates an objective ability of $P$.

\item An important mechanism is that $P$ feels more able (on performing tasks of class $X$)
by having been informed about $T$.%
\footnote{In this scenario it is plausible, but not necessary, that  group $A$ conveys a sense of self-presumed ability to $P$. This scenario represents the only justification for $A$ asking $P$ to take notice of $T$ in a situation where (i) a proof that $P$'s awareness of $T$ increases $P$'s ability  (on tasks of class $X$)
is not within reach, and, (ii)  obtaining majority support from the community for the assertion that awareness of $T$ should be counted as strengthening $P$'s competence is a premature ambition for author group $A$, and where (iii) mere self-endorsement (by $P$) is also considered too weak a basis for applauding $P$'s interest in $T$.}

Here $P$'s knowledge of $T$ generates an auto-subjective conjectural ability of $P$.

\item Finally, and most plausibly: the author group $A$ acts as a special interest group (preferably a group of
persons with community competence on $X$)  that maintains and advertises the
hypothesis that knowledge of $T$ will increase $P$'s ability concerning $X$ class tasks.%
\footnote{This state of affairs may apply simultaneously for different and mutually contradicting
theories $T$ and $T^{\prime}$.}

Here $P$'s knowledge of $T$ generates an exclusive intersubjective ability of $P$.
\end{enumerate}

\subsubsection{CCC+p(OA$_{ncc}$)+mCA$_{A,L}$ competence profile for $X$}
CCC abbreviates confirmed community competence, p(OA$_{ncc}$) abbreviates partial command of the
non community confirmed objective ability OA$_{ncc}$,  and CA$_L$ abbreviates the conjectural ability generated by knowledge of theory $L$. CCC results from
an observed competence acquisition process, pOA$_{ncc}$ results from reading some part of the
specialized scientific literature.
CA$_{A,L}$ results from taking notice of literature $L$ (concerning $X$) endorsed by a group A (probably including the authors of $L$, assuming that none of those operates with fraudulent intentions). p(CA$_{A,L}$) represents an ability as induced by a partial awareness of
$L$, where a filtering operator p selects a part of CA$_{A,L}$.
mCA$_{A,L}$ represents a combination of conjectural abilities
p$_1$(CA$_{A_1,L_1}$)+..+p$_n$(CA$_{A_n,L_n}$) for a number of theories $L_i$ and corresponding endorsing groups $A_i$.

The combination of CCC and one
or more of a family of CA's (each derived from its own strand of literature) provides a plausible description
of the state of competence, ability, and skill (together understood as a competence profile)
acquired by a person operating in a field where ideologies (often
called theories) lacking a convincing empirical evidence base exist besides a generally acknowledged
concept of community competence which is equally lacking a convincing evidence base in terms of its
proven functionality.

With ``$P$'s $X$-competence profile'' we will denote a person $P$'s state of competence,
ability (including skill), conjectural ability, concerning $X$.
Thus although abilities and skills are distinct from competences they are included in the
so-called competence profile (by lack of a better indication of a general category).

It seems that the S\&O field is likely to be inhabited by persons with a
CCC+p(OA$_{ncc}$)+p$_1$(CA$_{A_1,L_1}$)+..+p$_n$(CA$_{A_n,L_n}$)
 competence profile. This allows many combinations because many selections can be made for selecting
 pOA$_{ncc}$ from OA$_{ncc}$ and each theory $L_i$ concerning $X$ may induce a component of the profile,
 where each component p$_i$(CA$_{A_i,L_i}$) may consist of one of many selections from the knowledge items
 contained in $L_i$.

\subsection{Intentional classification of conjectural abilities}
Conjectural ability is an uncommon concept, and readers may be reluctant to trust it. Although a satisfactory
place for conjectural ability as a component of a competence profile has just been established, one may wonder where the boundaries of conjectural ability ought to be put.

The distinction between objective and subjective as well as the distinction between various kinds of subjective abilities, which has been phrased in terms of conjectural abilities has been made from the perspective of who
endorses the conjectural ability, not from the perspective of why that is done and with what intentions, or rather from the perspective of constraints imposed on working methods.

The methodological aspect of conjectural abilities can be approached by providing a further classification of conjectural ability in terms of the intentions of agents said to avail of such abilities. This classification
necessarily  includes also purported abilities that should be rejected on scientific grounds.

In the following classifications one imagines that a person $P$ has a (by default
self assessed) conjectural ability profile which consists of a range of conjectural abilities.

\begin{description}
\item{\em EDCA: Evidence disregarding conjectural ability.} EDCA is present (as an instance of exclusive intersubjective conjectural ability) if endorsing interest groups are
not worried by the presence of absence of evidence as gathered by the normal scientific research process.
Astrology may be a convincing representative of this category, as well as some forms of alternative medicine.%
\footnote{A specific EDCA (say for prognosis or healing) may be assigned to $P$ by
other persons even if $P$ denies any ground for that assumption. Such matters are not reflected in $P$'s conjectural ability profile.}
\item{\em EICA: Evidence immune conjectural ability.} The interest group endorsing an EICA sees to it that
it works with conjectures that are independent of evidence that can be collected by means of scientific research.%
\footnote{If one bases the ability to restructure society on conjectures regarding the intrinsic meaning of a particular religion, that can be done in such a way that the normal scientific process cannot either refute or confirm these assertions. The bundle of conjectures that underly a particular instance of EICA is often considered an ideology,
rather than a theory, but the difference is a matter of gradation. An EICA may range from auto-subjective and
exclusive intersubjective to community confirmed.}
\item{\em ERCA: Evidence respecting conjectural ability.} As soon as evidence emerges refuting a conjecture
that underlies the claimed conjectural ability the interest group endorsing it will change its view. That interest group
is uncommitted to obtaining either positive or negative evidence regarding the conjectures underlying the
claimed ability.

\item{\em EOCA: Evidence oriented conjectural ability.} EOCA is more restrictive than ERCA. In EOCA, the
 conjectures on which the claim of ability is
based are in principle amenable for scientific analysis, and sooner or later confirmation or refutation is expected
by those claiming the competence. In addition the belief is stated by the interest group backing this CA that confirmation will result, and moreover they demonstrate a keen interest in obtaining evidence in either way.%
\footnote{EOCA is likely to occur in medicine in cases where a field study provides a strong but inconclusive suggestion that some treatment works. Disallowing clients that treatment until the evidence base has been secured may not correspond to their own wishes. In these circumstances medical experts may be moved into
either an ERCA or an EOCA pattern of professional behavior, perhaps even against their own preferences.}

\item{\em EBCA: Evidence based conjectural ability.} This is what has been called ability before. The
phrase seems to be self-contradictory because in the presence of an evidence base for it an ability is not
conjectural anymore. We hold that the conjectural status can coexist with the knowledge of evidence.
This seems to be  a matter of psychology, rather than a matter of logic or of philosophy of science.%
\footnote{An unclear case arises if the evidence has been obtained by scientific fraud. It seems fair to hold that
the evidence for an ability base exists until appropriate
research journals have actually withdrawn their endorsement of
the papers that underly the claims of evidence, or until the protagonists
of the particular ability (unqualified) have withdrawn the statement that their claims can be based
on the results of said (now obsolete)
works in the case that no explicit withdrawal has taken place by its authors or by those responsible for its publication.}

If a mathematician $P$ conjectures the truth of mathematical assertion $\phi$ at some moment of time
and if $P$ is subsequently
notified at some later moment that $\phi$ has been confirmed by means of a rigorous proof by another mathematician
$P^{\prime}$,
must $P$ immediately thereafter  dismiss the conjectural  status of $\phi$ even if no details of the proof have yet been
disclosed (to $P$). Probably not, though on the basis of the reputation of $P^{\prime}$, $P$ may start believing the
 factual status of $\phi$ in addition. It is plausible to assume that for some time $P$ maintains some form of
 superposition of the conjectural and the factual status of $P$.
\end{description}

\subsubsection{Conjectural ability profile assessment}
Assuming that a person $P$'s conjectural ability profile (CA profile below)
is given, then how is that likely to be assessed by a candidate business partner $Q$ of $P$?
This question cannot be answered independently of the
underlying domain $X$. At this stage we  add the assumption
 that $X$ is outsourcing (or sourcing which makes little difference here).

If $P$'s CA profile
 claims any conjectural abilities classified under EDCA that is likely to be held against $P.$%
 \footnote{In some cultures EDCA type conjectural abilities are well-received (for instance when designing a new building) but there is no indication that this also applies to outsourcing (or sourcing for that matter).}
 Conjectural abilities classified as EICA are met with suspicion. If the underlying
ideology matches with that of $Q$, it will not always be appreciated if that ideology is made explicit by $P$
because doing so implicitly allows for the possibility that $P$ might not adhere to it.
If, however, the underlying ideology
differs from that of $Q$ a negative assessment by $Q$ of $P$'s conjectural ability profile is to be expected.%
\footnote{We assume that neither EDCA components nor EICA components of $P$'s CA profile can be constituents
of $P$'s community confirmed competence profile, because communities are unlikely to agree on the
plausibility of the underlying conjectures.}

ERCA components of $P$'s CA profile may well be appreciated by $Q$,
and an EOCA bias may be preferred assuming that the bias is credible. If there is no relevant ongoing research
and no clear perspective of getting it going, adopting EOCA classified conjectural abilities may be less credible
than adopting ERCA classified ones. Both ERCA components and EOCA component of a CA profile may be
constituents of $P$'s competence profile and in some cases also of $P$'s validated community competence.

The status of ERCA components of $P$'s CA profile is a non-trivial matter. If a component is community
confirmed that component is a constituent of $P$'s community confirmed competence profile and its presence in
$P$'s competence profile is likely to be applauded by $Q$. If, however, the ability $\alpha$ at hand is not community
confirmed, $Q$'s commitment to evidence based work is tested. $P$ needs to take into account that $Q$ may rather
follow its views concerning what warrants community confirmation (concerning the status of $\alpha$) than
to rely on an independent judgement of the evidence base available for $\alpha$.

The CA components that have been mentioned above as the part
p$_1$(CA$_{A_1,L_1}$) +..+p$_n$(CA$_{A_n,L_n}$) of a
CCC+p(OA$_{ncc}$)+p$_1$(CA$_{A_1,L_1}$) +..+p$_n$(CA$_{A_n,L_n}$) competence profile, and which result
from intended outsourcing theory development are most likely to be classified as ERCA components, in view of the circumstance that it is so hard to press for obtaining the evidence base required to
evaluate EOCA components. If $P$ lists an ERCA component $\alpha$ in his or her CA profile,  that by itself
need not contradict that $P$ maintains a
preference for EOCA components  over ERCA components. Neither does it reflect a lacking preference of $P$ for
EBCA components over EOBA components. It merely reflects $P$'s position towards the plausibility of collecting conclusive evidence in favor or against $\alpha$, together with $P$'s belief that working on the basis
of $\alpha$ will more effective than working with EBCA (conjectural) abilities or with EOCA conjectural abilities.

\subsubsection{Conjectural ability profile impact}
In most general terms the impact of a  competence profile, and in particular of the CA profile included in the competence profile, on a person's achievement on class $X$
tasks can be split in three parts, listed in the order of importance:
(i) functional  impact (that is impact on project outcomes),
(ii) project acquisition impact, and
(iii) project outcome marketing impact.

Part of the impact is only expected in a conjectural way,
and part of the impact may be expected on the basis of evidence.
We formulate a hypothesis on the impact of an  outsourcing  conjectural ability p(CA$_L$) resulting
from awareness of theory and or of empirical research results on outsourcing.
This conjectural ability  has most (expected) impact during early stages of the life-cycle of an sourcing transformation project, while in later stages of a sourcing transformation project specific details as well as general skills and competences become more important for the project.

\subsection{S\&O ability (unqualified) acquisition}
Acquisition of S\&O abilities amounts to acquisition of conjectural abilities
as in the absence of recognized objective abilities specific for S\&O there are no community confirmed objective
abilities in that area either.

The limitations of this matter are connected with the lack of convincing meta-studies that combine empirical results from different groups into a uniform framework.%
\footnote{Having performed research in an S\&O area usually creates corresponding literacy as well as an
awareness of how difficult it is to obtain reliable and reproducible data concerning S\&O which exceeds
the anecdotical level.}

\subsubsection{Community confirmed conjectural abilities}
It is reasonable to assume that increased S\&O community confirmed competence
correlates with increased (community confirmed) conjectural S\&O ability. The presence of this
correlation is only plausible, however, if the S\&O competence at hand has been acquired in an efficient manner.
Otherwise it might have been better for the person involved to acquire other  competences which are in more demand
instead.

Community confirmed conjectural abilities, are generated by (that is conjectured on the basis of)
community confirmed competences. Each class of community confirmed competences, obtained by an appropriate instantiation of the generic competence scheme, may generate conjectural abilities. This can in principle be worked
out in detail but we provide some examples only.
\begin{itemize}
\item Having read extensively about research on EU procurement both for service acquisition and in
cases of sourcing transformations creates theoretical literacy on ``EU procurement and outsourcing''. That in turn
leads to a conjectural ability to handle difficult EU procurement issues correctly (though not necessarily efficiently).
\item Having dealt with EU procurement in the setting of outsourcing in several domains creates general ``EU procurement and outsourcing'' competence which leads to a conjectural ability to deal with complex
cases of that kind effectively.
\item Theoretical literacy competence on outsourcing acquired by reading well-known works on
outsourcing and other sourcing transformations creates  conjectural ability to deliver meaningful courses on outsourcing.
\end{itemize}

\subsubsection{Exclusive intersubjective conjectural ability}
A second source of competence is related to knowledge that is produced for its own sake and
not as a byproduct of practice, and which is communicated as such via various communication
mechanisms. This knowledge is often consumed by a small group of stake-holders only. That renders it
non-community confirmed conjectural ability. We notice:

\begin{enumerate}
\item
We assume that non-community confirmed conjectural  S\&O ability will  increase with  increasing
theory development competence and with increasing empirical research competence.
Several caveats are in place, however:
\begin{itemize}
\item
Foundational theory development competence may have become high at the cost of other  highly regarded S\&O competences.
\item
This caveat even holds in the case of literature literacy on S\&O which may also be costly to acquire.
\item
Theory development competence (and more generally S\&O research competence) is assumed to produce an increase of conjectural
S\&O ability which is not in excess of the increase which results from acquiring the level of S\&O literacy that is implied by performing the mentioned theoretical S\&O research.\footnote{In other words: participation in research on S\&O increases one's S\&O competence no more than it would be increased by taking good notice of the research outcomes and of the works on which the research is based.}
\end{itemize}

\item Only with a theory at hand one can be specific about the conjectural abilities that awareness of that theory
may give rise to. In principle at this place in the argument a survey of non community confirmed theories on outsourcing might
be provided together with a suggestion of what conjectural abilities these may imply. In fact, however, we are not
aware of any theory of outsourcing that might be dealt with in that way.

\item In principle one may first specify an emergent ability, or in other words a candidate conjectural ability, (say
regarding outsourcing) and then ask for the development of a theory the availability of which plausibly leads to generating that conjectural ability (that is promoting it from an emergent status to a conjectural status).

\end{enumerate}

\section{Options for outsourcing theories}\label{Options}
There is no such thing as a single or most plausible theory of outsourcing. In the following Section we will propose requirements on a theory of outsourcing corresponding to our own preferences.
A complication that needs to be confronted first is that theory development can be approached in different ways.
Choosing a methodology for outsourcing theory development itself can be contemplated in the context of community
confirmation. Assuming that an outsourcing community considers outsourcing research, will it spot any method
of theory development as most plausible for the subject? When considering recent literature on outsourcing it seems to be the case that the 45 years old paradigm of grounded theory is becoming prominent in the area of management  and outsourcing. Some references for grounded theory are \cite{Turner1981} and \cite{Parry1998}. Examples of putting
outsourcing theory development in the perspective of grounded theory is given in \cite{Willekens2007} and
\cite{Kreeger2007}. The latter paper indicates that the paradigm of grounded theory facilitates constructionist ambitions
which are less appreciated by common positivist approaches to the social sciences.

In \cite{ChristensenSundahl2001} one finds a survey of approaches to theory formation in management science
suggesting that grounded theory may be applicable to outsourcing. The paper emphasizes the role of anomalies and the
need to look systematically for phenomena that are inconsistent with the current (grounded) theory.

A clear example of a fragment of outsourcing theory that comes about in a grounded fashion is found
in \cite{PerronsPlatts2004} where units are classified according to their so-called clock speed, that is an intrinsic rate
of innovation, and the conclusion is drawn that long term outsourcing relationships are more likely to be useful
for units with medium clock speeds. Assuming that most units under government control are of that nature the EU procurement regime is less plausible for such units because it disallows striving for long term outsourcing relationships.

Grounded theory advances social sciences by inferring theoretical models from empirical data in such a way that the
gap between theory and evidence is kept limited. Grounded theory may even hold the claim that theory must be grounded
in empirical data to find any legitimacy at all. Grounded theory has been developed as a tool for social sciences, especially
for dealing with unstructured data from free format interviews. Gathering such data is promoted because strictly formalized interviews may hide important qualitative aspects from researchers' attention. Given the importance of interview based
longitudinal research on heterogenous collections of case studies grounded theory may be well-suited as a methodology
for analyzing individual behavior in the context of outsourcing tasks. Grounded theory is especially helpful for persons or teams who are directly involved in the process of gathering yet uninterpreted data.

The kind of theory that we have in mind, originates at a larger distance from systematically collected empirical data and
results from a conceptual analysis, than seems plausible for a grounded theory developed during the process of data collection. A grounded theory may provide illuminating insights in a collection of otherwise unstructured data, but it
is unlikely to reach much further than available data into speculation about circumstances that have not yet been encountered.

\section{Requirements on a theory of outsourcing}\label{ReqToS}
We will now turn the issue around. Suppose a theory of outsourcing is to be developed. Then,
 (i) we must be clear about
what kind of questions must be dealt with by the intended theory, (ii) how it leads to an extension of
a person $P$'s competence profile of his confirmed community competence profile, and
(iii) it must be specified somehow,
(preferably per fragment of the theory) to which conjectural ability that may plausibly give rise.

Here is a listing of issues that a theory of outsourcing should deal with. Formulating corresponding conjectural abilities can only be done once an outsourcing theory has been laid down in sufficient detail.
\begin{description}

\item{\em Stratified approach: three level terminology.}
The subject should
be approached in a stratified fashion which allows three levels of static (or equilibrium)
aspects (descriptions, specifications, perspectives) and for each of these a corresponding transformational level. The three levels we have in mind  are:
\begin{itemize}
\item facts on the ground,
\item business models and cases,
\item contracts.
\end{itemize}

Stratification means here that facts on the ground must not be defined in terms of business models
or contracts and that business models and business cases may be explained on the basis of the facts on the ground but not in terms of contracts.

By making use of a stratification of concepts circular
dependencies between them will be prevented.  Taking housing as an example: if $A$ inhabits home $H$ owned by $B$ (factual level information), then more specifically $A$ may be renting $H$ from $B$ (business level information). That fact may be regulated by means of some contract $C$ (contract level information). $A$ may expect to leave $H$ (transformational level information) at time $t$ because this regulation occurs in $C$.

We notice that the assertion that $A$ is renting $H$ makes sense because $A$
inhabits $H$, (not because of the existence of a contract which then would lead to a circularity) whereas how to obtain modifications of the renting process is regulated by  contract $C$. Neither the contract, nor the phenomenon of renting are needed, however, to grasp the facts regarding $A$ living in $H$ and of $H$ being in the possession of $B$.

Avoiding circularities which may render large fragments of theory essentially
meaningless strikes us as being most important.
Removing circularities often requires the introduction of new or unusual aspects which support a hierarchical configuration that was absent or invisible before. Thus the focus of the theory will be on non-circular definitions and specifications. The key ingredient for that is to develop a hierarchically layered configuration for the relevant concepts together with a terminology that allows one to be precise about aspects relevant to a certain layer without even mentioning any aspects of higher layers.

Consider the  issue about $A$ renting from $B$ raised above. In this example three levels of description can be disentangled and non-circular definitions of the underlying concepts can be given. Can equally clear distinctions be found in the subject of sourcing thus leading to non-circular concept descriptions
in a similar fashion?%
\footnote{In \cite{Thompson1956} an observation is made that might be read as an indication that the three level
stratification just proposed is implausible. In order to infer business model from the facts on the ground one must
perform some form of abstraction. This can only be done if some form of business context is given
as a point of departure. It may be the case that contracts need to be read in order to decide how to determine
business models fro the facts. This objection is important and it leads to the conclusion that a business model
can be presented on top of the facts without any mention of contracts but it may be impossible to infer the model without access to the contractual configuration.}

\item{\em Explanation of the concept of source.} A theory of outsourcing must pay due attention to the
notion of source. We provide some preliminary remarks here:

\begin{description}
\item{\em What is a source?}
Units (or organizational units) make use of sources. A very characteristic example of a
source is a computer network on a university campus where the university is seen as a unit. The
remarkable strength of the term source transpires if one tries to formulate what else the
unit may be about: for what are these sources used? The term mission is too abstract, the phrase core competence is too instrumental, the term capability comes close but is still too abstract.
We suggest to assume that a unit $U$
runs one or more primary processes which
(i) embody  the realization of its mission,
(ii) materialize the realization of its capabilities, and
(iii) capture its performance in accordance with its strategy.

These primary processes may in turn constitute
sources (in the form of services offered by the unit running the processes) which are
made use of by one or more different units but from the perspective of $U$ the
inputs, outputs, and flows of its primary processes do not constitute sources.

\item{\em Transferability of sources.}
Sources can be owned, acquired, bought, sold, stolen or given away, leased, hired,
rented, designed, developed, invented, patented, build, constructed, removed, used, or left idle.
An attribute of a source or of any part of a primary process is not a source by definition.
Thus, quality, quantity, color, age, size, number, any KPI, or a role like client or server, do not
qualify as sources. To find out what sources a unit is using one may need to design
an ontology for it, then to determine its primary processes and now to skip all terms
denoting parts of primary processes and all terms serving as attributes or qualifications
of something else.

\item{\em Relativity of sources.}
The intuition of a source is not at all an obvious one. Consider a museum $M$ which
owns a large collection of items most of which are not exposed to the public at any
specific moment of time. If $M$ has  a strong identity, which is independent of the works
of art
it exposes, its entire collection of items may be considered a source, or each of the items
may be considered individual sources. Alternatively, however, the mission statement
of a museum may include the permanent exposition of one or more works of art
(for instance: Vermeer's Melkmeisje in the Mauritshuis in The Hague, or Rembrandt's
Nachtwacht in the Rijksmuseum in Amsterdam). In such a case
it is unconvincing to regard
these crucial items as sources because they enter the description of the mission statement
of the museum and for that reason they constitute part of one of the museum's
primary processes. It follows that the concept of a source is relative to the mission of a unit.

In \cite{BDV2011a} the qualification of sources as being mission defining or mission co-defining
has been discussed. One might consider Vermeer's Melkmeisje a mission co-defining source
for the Mauritshuis. But to consider the Eiffel tower a mission defining source for the unit that
exploits its access for visitors is hardly convincing. This remarkable
artifact has the visibility, presence and strength to bring about
a unit with the corresponding mission statement if it were missing at any time.

\item{\em Sources in the context of IT sourcing.}
In the special case of IT-sourcing the conceptual problem of what constitutes a source
is absent if IT is merely used as a tool. Thus for a museum that is not holding computers and
software itself as items in its collection, all aspects of its IT are related to sources. The
same may hold for a school. But for a training center of a software company some of its
own software will not have the status of source but instead it will be considered a constituent
of its primary process. IT-sourcing then concerns only part of the company's IT assets.
\end{description}

\item{\em Modularity.}
A fine-grained theory of outsourcing supports a modular view. That in turn depends on clarity about basic building blocks. A theory of outsourcing must support a modular approach to the topic.

A module is a component in a context where components can be combined. sourcements describe fragments of a reality  and these descriptions can be combined by taking their union. Sourcements will play the role of modules in a modular theory of sourcing. Admittedly this is a rather weak form of
combination but it serves our purposes. In a modular theory of sourcing all aspects of sourcing
and sourcing transformations are related to modules (in this case sourcements) that feature only the aspects needed for that particular aspect.

\item{\em Follow-up outsourcing.}  As time passes the share of initial outsourcing transformations (as a fraction of all sourcing transformations) will decrease and the fraction of follow-up outsourcing transformations will increase. In a follow-up outsourcing transformation sources may move to a new (other) unit once more. A precise definition of follow-up outsourcing must be given, and also of backsourcing which is a special case of follow-up sourcing.

\item{\em Transformation and transition description.}  Once a sourcing transformation has been specified its implementation takes place by putting a transition into effect. An outsourcing theory must provide tools for
the notation of algorithms (instruction sequences) or more abstractly control codes
(see \cite{BergstraMiddelburg2009} and \cite{Bergstra2010b}) which when put into effect
have the required results. Both risk analysis and testing surface as vital elements in this stage. The discussion of risks and testing in \cite{Bergstra2010b} applies to control codes for  sourcing transformations as well.

\item{\em Transformation independent business case.} For some sourcing transformations it may be unclear to what extent the jargon of outsourcing and insourcing is adequate. Then it must be possible to accumulate information in more neutral terms before deciding to label an expected transformation with familiar but only seemingly specific labels, such as outsourcing. Because \cite{Delen2005} claims that both parties knowing both business cases is a critical success factor for outsourcing, it may be useful to have a terminology available where the business case is analyzed while avoiding the label outsourcing so that another next step can be envisaged making use of the same business case information. Strictly speaking these business cases concern new sourcements rather than the economics of the transformation into these new sourcements.

\item{\em European Tender Procedure (ETP).}
European regulations add a significant dimension of complexity to the concepts of sourcing and outsourcing.
Without paying due attention to the matter, the development of a theory of  (out)sourcing
leads to mismatches with the jargon and processes that are being enforced by the EU  by way of the European Tender Procedure (ETP) see \cite{ETPb} and \cite{ETPa}.%
\footnote{Instead of ETP one often speaks of EU procurement, sometimes of European procurement. In
\cite{RoodhooftVdA2006} PTP is used for European Public Tendering Process. ETP is  termed  the EU public procurement regime in \cite{NielsenHansen2001}. }

Literature about outsourcing in connection with EU procurement is not easy to find. Most EU procurement literature concerning services approaches the matter as a unidirectional purchase of a service contract. Nevertheless besides
conceptual works relating outsourcing and ETP the study of ETP's effectiveness for outsourcing from public units
is important. We refer to \cite{NielsenHansen2001} for an analysis of the economic principles of such work.

Outsourcing theory must be developed in such a way that the relation with ETP is clarified.%
\footnote{The impact of ETP can be quite specific for a kind of business.
 In the Netherlands, for instance, a substantial amount of work has been done on the
impact of EU procurement in the domain of home care (e.g. see \cite{Staveren2010}). In that domain it is far from obvious to decide when ETP is mandatory and ETP has many unexpected and costly side-effects. With no sign of service offerings from outside The Netherlands emerging in this domain, EU regulations have had a counterproductive impact  on the procurement
and provision of care services by frustrating communication between different parties who ought to
communicate in order to best serve their clients.}

\item{\em Normal case.} Units may be classified by means of a type. Then different units of the same type can be compared. Having such information available one may find that for certain sources in certain units in certain environments it is normal (majority of cases) that these units contain those sources.

\hspace{5mm}In the Platform Outsourcing Nederland (PON) standard outsourcing contract \cite{PON2006} one seems to need the following interpretation of outsourcing: ``having been outsourced in the light of the fact that
this is not normally the case''. Clearly this usage of the term outsourcing is at odds with the convention that
we have opted for in \cite{BV2010}.

\hspace{5mm}This raises the important question if normal-case considerations must enter outsourcing theory. We have no opinion about this matter, but it needs to be covered by an outsourcing theory.

\item{\em Just in time sourcing method selection.} When an outsourcing is envisaged it may not yet be known to what extent there will be a need or an opportunity to split the sourcement envisaged for outsourcing into different lots in order to make use of the competences and workforce of more than a single provider. A flexible language is needed which accommodates the possibility that such information may become available only at some specific stage in a formalized transformation process. Moreover, that allows that the choice of such a process itself, at some stage, has not yet been made. In other words, that making the choice is itself a part of the formalized process.

\end{description}

\section{Concluding remarks}\label{Conclusions}
In this paper the results of \cite{BV2010} and \cite{BDV2011a} have been used and extended towards
an analysis of sourcing and outsourcing competence. It has been argued that outsourcing competence
can stand on its own feet and has a role to play besides domain specific sourcing competences and besides a range of service industry competences.

Outsourcing competence as its stands today need not be understood as being evidence based, it
may be better characterized as a so-called community competence. The notion of ability has been positioned to indicate the higher level interpretation of competence which will be evidence based at least in principle. So, at the moment outsourcing competence will not guarantee outsourcing ability, and
in fact assessing a person's outsourcing ability objectively is hardly possible. In order to deal with the problem that
an (objective) ability more often than not can't be confirmed either, (which is plausible in the subject of outsourcing), in the near future, the notion of conjectural ability has been put forward.

Further we have suggested that outsourcing theory besides being of an independent interest as
a topic of pure investigation, should be understood as a tool for strengthening an outsourcing
competence profile by augmenting it with  specific conjectural abilities. That awareness of some theory
of outsourcing creates a plausible conjectural ability can be understood as a requirement from
a practical perspective on
the development of an outsourcing theory. Based on that view a survey of issues that outsourcing theory should eventually cover has been given. This suggests that a competence profile pull, and more specifically a conjectural ability pull, rather than a theory push can and preferably should be driving the development of outsourcing theory.
Following this line of thought the development of an outsourcing theory can be motivated as working towards a way to offer a new or improved conjectural ability component of a person's  outsourcing competence profile.


\begin{thebibliography}{99}
\bibitem{Bergstra2010b}
J.A.\ Bergstra.
\newblock Informal Control Code Logic.
\newblock 2010.
\newblock {\tt arXiv:1009.2902 [cs.SE]}.

\bibitem{BDV2011a}
J.A.\ Bergstra, G.P.A.J.\ Delen, and S.F.M.\ van Vlijmen.
\newblock Introducing Sourcements.
\newblock {(\tt arXiv:1107.4684 [cs.SE])}, (2011).

\bibitem{BergstraMiddelburg2007}
J.A.\ Bergstra and C.A.\ Middelburg.
\newblock Thread algebra for strategic interleaving.
\newblock {\em Formal Aspects of Computing}, 19 (4):445--474, (2007).

\bibitem{BergstraMiddelburg2009}
J.A.\ Bergstra and C.A.\ Middelburg.
\newblock Machine structure oriented control code logic.
\newblock {\em Acta Informatica}, 5(1):170--192, (2009).
\newblock {(\tt arXiv:0711.0836 [cs.SE])}.

\bibitem{BergstraMiddelburg2011}
J.A.\ Bergstra and C.A.\ Middelburg.
\newblock An application specific logic for interest prohibition theory.
\newblock {(\tt arXiv:1104.0308 [q-fin.GN])}, (2011).

\bibitem{BV2010}
J.A.\ Bergstra and S.F.M.\ van Vlijmen.
\newblock Business mereology, imaginative definitions of outsourcing and insourcing transformations.
\newblock  {\tt arXiv:1012.5739 [cs.SE]}, (2010).

\bibitem{Beulen2000}
E.P.\ Beulen.
\newblock Beheersing van IT-outsourcingrelaties.
\newblock {\em PhD Thesis, University of Tilburg},  in Dutch, (2010).

\bibitem{Burgess2007}
M. Burgess.
\newblock System administration and the scientific method.
\newblock in: J.A.Bergstra and M. Burgess (editors),
{\em Handbook of Network and
System administration}:
\newblock pp.\ 689-728, (2007).

\bibitem{ChristensenSundahl2001}
C.M.\ Christensen and D.M.\ Sundahl.
\newblock The process of building theory.
\newblock Harvard Business School, Boston MA, Working Paper 02-016 (2001).

\bibitem{CurtolPB2006}
F.\ Curtol, G.\ Peserin and T.\ Vander Beken,
\newblock Testing the mechanism on EU procurement legislation.
\newblock {\em Euopean Journal of Criminal Policy Research}, Vol 12, pp. 337-364 (2006).

\bibitem{Delen2005}
G.P.A.J.\ Delen.
\newblock Decision- en controlfactoren voor sourcing van IT.
\newblock {\em PhD Thesis, University of Amsterdam}, Van Haren Publishing Zaltbommel, in Dutch, (2005).

\bibitem{Delen2007}
G.P.A.J.\ Delen.
\newblock Decision and Control Factors for IT-sourcing.
\newblock in: J. A.Bergstra and M. Burgess (editors), {\em Handbook of Network and System administration}:
\newblock pp.\ 929-946, (2007).

\bibitem{ETPa}
European Tender Procedure,
\newblock http://ec.europa.eu/youreurope/business/ \\profiting-from-eu-market/benefiting-from-public-contracts/index\_en.htm.

\bibitem{ETPb}
\newblock Official Journal L 134, 30.4.2002, p. 1140240. (2002).

\bibitem{KernWillcocks1999}
T.\ Kern and L.P.\ Willcocks.
\newblock Exploring information technology outsourcing relationships: theory and practice.
\newblock {\em Erasmus University Rotterdam}, management report no. 61-1999, (1999).

\bibitem{KimblerBouma1995}
K.\ Kimbler, L.G.\ Bouma (Eds.)
\newblock Feature interactions in telecommunications and software systems V.
\newblock {IOS Press}, (1995).

\bibitem{Kreeger2007}
L.D.\ Kreeger.
\newblock Inside outsourcing: a grounded theory of relationship formation within a nascent service system.
\newblock Antioch University, Ph D. Thesis (2007).

\bibitem{Lacity1993}
M.C.\ Lacity.
\newblock Information systems outsourcing: myths, methaphors and realities.
\newblock {\em Wiley, Chishester}, (1993).

\bibitem{LohVenkatraman1992}
L.\ Loh and N.\ Venkatraman.
\newblock Diffusion of information technology outsourcing, influence sources and the Kodak effect.
\newblock {\em Information Systems research}, 4, pp 334-358, (1992).

\bibitem{MPBR2002}
J.\ Mills, K.\ Platts, M.\ Bourne and H.\ Richards.
\newblock{\em Strategy and performance, competing through competences.}
\newblock{Cambridge University Press}, (2002).

\bibitem{NielsenHansen2001}
J.U.\ Nielsen and L.G.\ Hansen.
\newblock The EU public procurement regime-does it work?
\newblock {\em Intereconomics.} Vol. 36 (5) pp. 255-263 (2001).

\bibitem{Parry1998}
K.\ Parry.
\newblock Grounded theory and social process: a new direction for leadership research.
\newblock{\em Leadership Quarterly,} Vol. 9 (1) pp. 85-105 (1998).

\bibitem{PerronsPlatts2004}
R.K.\ Perrons and K.\ Platts.
\newblock The role of clockspeed in outsourcing decisions for new technologies:
insights from the prisoner's dilemma.
\newblock{\em Industrial Management and Data Systems} Vol. 104 (7) pp. 624-632 (2004).

\bibitem{PfefferSutton2006}
J.\ Pfeffer and R.I.\ Sutton.
\newblock Evidence-based management.
\newblock {\em Harvard Business Review, OnPoint Article}, Product 298X, (2006).

\bibitem{PON2006}
Platform Outsourcing Nederland, werkgroep Taxonomie Outsourcing.
\newblock Outsourcing van IT -- Management guide.
\newblock Van Haren Publishing, (2006).

\bibitem{RobothamJubb1996}
D.\ Robotham and R. Jubb.
\newblock Competences: measuring the unmeasurable.
\newblock {\em Management Development Review}, Vol 9 (5), pp. 25-29 (1996).

\bibitem{RoodhooftVdA2006}
F.\ Roodhooft and A.\ van den Abbeele.
\newblock Public procurement of consulting services.
\newblock {\em International Journal of Public Sector Management,} Vol. 19 (5) pp. 400-512 , (2006).

\bibitem{Schettkat2010}
R.\ Schettkat.
\newblock Something unforeseeable happened? National rate theory and economic crisis.
\newblock {\em Intereconomics,} Vol. 45 (5) pp. 297-304 (2010).

\bibitem{Shippmann2000}
J.S.\ Shippmann et. al.
\newblock The practice of competency modeling.
\newblock {\em Personnel Psychology}, 53 pp. 703-739, (2000).

\bibitem{Staveren2010}
I. van Staveren.
\newblock Home care reform in the Netherlands: impact on unpaid care in Rotterdam.
\newblock {\em Forum Social Economics,} 39: pp. 13-21, (2010).

\bibitem{Tampoe1994}
M.\ Tampoe.
\newblock Exploiting the core competences of your organization.
\newblock {\em Long Range Planning,} Vol. 27 (4) pp. 66-77 (1994).

\bibitem{Thompson1956}
R.H.\ Thompson.
\newblock The subjective element in archeological inference.
\newblock {\em Southwestern Journal of Anthropology,} Vol. 12 (3) pp. 327-332 (1956).

\bibitem{Trout2002}
J.D.\ Trout.
\newblock Scientific explanation and the sense of understanding.
\newblock {\em Philosophy of Science,} Vol. 69 (2) pp. 212-233 (2002).

\bibitem{Turner1981}
B.A.\ Turner.
\newblock Some practical aspects of qualitative data analysis: one way of organizing
the cognitive process associated with the generation of grounded theory.
\newblock{\em Quality and Quantity,} Vol. 15, pp. 225-247 (1981).

\bibitem{VargoLusch2004}
S.\ Vargo and R.F.\ Lusch.
\newblock Evolving to a new dominant logic for marketing.
\newblock {\em Journal of Marketing} 68 pp.\ 1-17, (2004).

\bibitem{Waller2001}
B.N.\ Waller.
\newblock Classifying and Analyzing Analogies.
\newblock {\em Informal Logic,} Vol.\ 21 (3), pp.\ 199-218, (2001).

\bibitem{WaltonMacagno2010}
D.\ Walton and F.\ Macagno.
\newblock Defeasible classifications and inferences from definitions.
\newblock {\em Informal Logic,} Vol.\ 30 (1) pp.\ 34-61, (2010).

\bibitem{Willekens2007}
M.\ Willekens.
\newblock Improving offshore outsourcing success with Carnegie Mellon's
eSourcing Capability Model.
\newblock MSc Thesis on technology management, Groningen State University (2007).

\end{thebibliography}
\end{document}